\documentclass[11pt,a4paper]{article}
\pdfoutput=1
\usepackage{putex}
\usepackage{bm}
\usepackage{soul}
\usepackage{cite}
\usepackage{tikz}
\usepackage{graphicx}
\usepackage{hyperref}
\usepackage{mathrsfs} 
\usepackage[utf8]{inputenc}
\usepackage{amsmath,amssymb}
\newcommand{\taubar}{\overline{\tau}}
\newcommand{\qbar}{\overline{q}}

\newcommand{\Sp}{\mathrm{Sp}}
\newcommand{\SL}{\mathrm{SL}}
\newcommand{\PSL}{\mathrm{PSL}}
\newcommand{\GL}{\mathrm{GL}}
\newcommand{\rmO}{\mathrm{O}}
\newcommand{\U}{\mathrm{U}}

\newcommand{\imt}{\textrm{Im}\tau}
\newcommand{\zbar}{\bar{z}}
\newcommand{\grp}{\Lambda^*/\Lambda}
\newcommand{\dt}{|\textrm{det }Q|^{-\frac{1}{2}}}

\newcommand{\ds}{\displaystyle}

\newcommand{\Z}{\mathbb{Z}}

\begin{document}

\title{Chern-Simons Invariants from Ensemble Averages}

\authors{\Large Meer Ashwinkumar$^{\bullet}$, Matthew Dodelson$^{\bullet}$,  Abhiram Kidambi$^{\bullet}$,\\[10pt]  Jacob M.\ Leedom$^{\ddagger, \ast}$ and Masahito Yamazaki$^{\bullet}$}

\date{}

\institution{}{\centerline{$^{\bullet}$Kavli IPMU, University of Tokyo, Kashiwa, Chiba 277-8583, Japan}}
\institution{}{\centerline{$^{\ddagger}$Theory Group, Lawrence Berkeley National Laboratory, Berkeley, CA 94720, USA}}
\institution{}{\centerline{$^{\ast}$Berkeley Center for Theoretical Physics, University of California, Berkeley, CA 94720, USA}}

\abstract{We discuss ensemble averages of two-dimensional conformal field theories associated with an arbitrary indefinite lattice with integral quadratic form $Q$. We provide evidence that the holographic dual after the ensemble average is the three-dimensional Abelian Chern-Simons theory with kinetic term determined by $Q$. The resulting partition function can be written as a modular form, expressed as a sum over the partition functions of Chern-Simons theories on lens spaces. For odd lattices, the dual bulk theory is a spin Chern-Simons theory, and we identify several novel phenomena in this case. We also discuss the holographic duality prior to averaging in terms of Maxwell-Chern-Simons theories.}


\pagenumbering{Alph} 
\maketitle  
\pagenumbering{arabic} 
\tableofcontents  

\section{Introduction}

One of the most remarkable recent developments in quantum gravity is the realization that semi-classical Euclidean quantum gravity requires us to sum over \emph{ensembles} of semi-classical geometries, at least for two-dimensional Jackiw-Teitelboim gravity \cite{Saad:2019lba,Stanford:2019vob} and three-dimensional pure gravity \cite{Maxfield:2020ale}. 
In the language of holography, such an ensemble average in the bulk is translated into an ensemble of conformal field theories (CFTs). It is therefore of great interest to further study ensembles for a simple class of CFTs, and discuss their holographic interpretations. This will hopefully shed light on the question of when and how ensemble averages arise more generally in holography.
For non-supersymmetric CFTs we generically do not expect any moduli space. However, there are still known examples of moduli spaces of CFTs without supersymmetry in the literature \cite{Narain:1985jj,Narain:1986am}. In the recent references \cite{Afkhami-Jeddi:2020ezh,Maloney:2020nni}, the ensemble average of toroidally-compactified free bosons has been considered.
The moduli space in this case is the Narain moduli space:
\begin{align}
\mathcal{M}_{\mathrm{II}_{p,p}}=\rmO(p,p;\mathbb{Z}) \big\backslash \rmO(p,p; \mathbb{R}) \big/(\rmO(p; \mathbb{R})\times \rmO(p; \mathbb{R})) \;,
\label{Mpp}
\end{align}
where $\mathrm{II}_{p,p}$ denotes the even, self-dual lattice associated with the
compactification on the $p$-dimensional torus $\mathbb{T}^p$.
The resulting average was then interpreted in the holographic dual as an exotic gravity theory approximated by the
Abelian Chern-Simons theory with gauge group $\U(1)^{2p}$. This holographic duality was studied further in e.g.\ \cite{Perez:2020klz, Dymarsky:2020bps,Dymarsky:2020pzc,Meruliya:2021utr,Datta:2021ftn,Benjamin:2021wzr,Meruliya:2021lul}.
In this paper we consider the generalization where the associated CFT moduli space is
a more general type of Narain moduli space
associated with an indefinite quadratic form $Q$ of rank $p+q$ and signature $(p,q)$\footnote{We will also denote $p-q$ as the signature in some parts of this paper. The usage is clear from context.}:
\begin{align}
\mathcal{M}_{Q} =
\mathrm{O}_Q(\mathbb{Z}) \big\backslash (\rmO(p,q; \mathbb{R})\big/(\rmO(p; \mathbb{R})\times \rmO(q; \mathbb{R})) \;,
\label{Mpq}
\end{align}
where $\rmO_Q(\mathbb{Z})$ is a subgroup of $\mathrm{O}(p, q;\mathbb{Z})$ preserving the quadratic form $Q$. 
The dimension of this moduli space is
\begin{align}
\textrm{dim}_{\mathbb{R}}\, \mathcal{M}_Q =  p q\;.
\end{align}
In the process of generalizing to an arbitrary integral quadratic form $Q$, we will encounter many interesting 
features which were not present in the previous studies.
Our discussion applies to non-self-dual lattices, and additionally to lattices with $p\ne q$ (such as those arising from toroidal compactifications of the heterotic string theories), where we have gravitational anomalies.
Finally, we are also able to analyze odd integral lattices, where
the partition function is dependent on the choice of spin structure.
Our analysis shows that the partition function after the ensemble average
contains spin Chern-Simons invariants for the handlebody geometries, 
giving further support to the appearance of the Chern-Simons term in the holographic dual.

The rest of this paper is organized as follows.
In section \ref{sec.bosonic}, we discuss the case where the integral quadratic form $Q$ is even.
We find that the ensemble average of the CFT partition function is equal to an Eisenstein series associated with $Q$, which can be interpreted as a sum over geometries in the three-dimensional Chern-Simons theories.
In section \ref{sec.fermionic}, we extend the discussion to an odd integral quadratic form.
In this case we have a non-trivial dependence on the choice of spin structure, and we identify the holographic dual to be a spin Chern-Simons theory. While the discussions in sections \ref{sec.bosonic} and \ref{sec.fermionic} address the holographic duality after ensemble averaging,
in section \ref{sec.before} we discuss holography before ensemble averages.
Finally, section \ref{sec.discussion} is devoted to a summary and concluding remarks.
We include appendices on technical materials.

\section{Ensemble Average of Bosonic CFTs}\label{sec.bosonic}

\subsection{Lattices with Even Quadratic Forms}\label{sec.lattice}

In this section we consider free boson CFTs with momenta valued in a $(p+q)$-dimensional integral lattice $\Lambda=\mathbb{Z}^{p+q}\subset \mathbb{
R}^{p+q}$,
equipped with an even quadratic form
\begin{align}
Q(\ell)=\sum_{i,j=1}^{p+q} Q_{ij}\ell^i \ell^j
\end{align}
with signature $(p, q)$. This quadratic form is said to be even if the value $Q(\ell)$ is even for any integral vector $\ell^i$, of length $p+q$.
This condition implies that $Q_{ii}$ is even for any $i$,
and $Q_{ij}$ is an integer for $i\ne j$;
it also implies the integrality of the bilinear form
\begin{align}
Q(\ell,m):=\frac{Q(\ell+m)-Q(\ell)-Q(m)}{2}=\sum_{i,j=1}^{p+q} Q_{ij}\ell^i m^j \;.
\end{align}

Before coming to the discussion of general $Q$,
it is useful to remind ourselves of the simplest case of the
$S^1$-compactification of the free boson. In this case we have $p=q=1$,
and the lattice $\Lambda$ is given by
\begin{align}
\Lambda=\left\{ \left(p^L=\frac{n}{2R}+ w R, \,\, p^R=\frac{n}{2R}- w R \right) \in \mathbb{R}^{2} \Big| \, n, w\in \mathbb{Z} \right\} \;,
\end{align}
where the integers $n$ and $w$ represent the momentum and winding, respectively.
The radius $R$ of the circle is the coordinate for the Narain moduli space. The quadratic form for this example is
\begin{align}
Q (\{n,w\})=2 n w=p_L^2-p_R^2   \in 2 \mathbb{Z} \;,
\end{align}
which determines the so-called $\mathrm{II}_{1,1}$ lattice.
Note that $Q$ is independent of the modulus $R$, while the choice of the modulus
is equivalent to the choice of the decomposition of the quadratic form $Q$ into
two positive definite quadratic forms $Q_L:=p_L^2$ and $Q_R:=p_R^2$
defined on one-dimensional subspaces $V_L, V_R$ of $\mathbb{R}^2$.
Such a choice is also equivalent with the choice of a positive quadratic form
on the whole of $\mathbb{R}^2$:
\begin{align}
H(\{ n, w\}):=Q_L+Q_R=p_L^2+p_R^2=\frac{n^2}{2R^2}+ 2w^2 R^2 \;.
\end{align}

Let us now discuss the case of a general even quadratic form $Q$.
The point of the moduli space $\mathcal{M}_{Q}$ is again specified by
decomposing the quadratic form into left and right-moving parts $Q_L$ and $Q_R$.
Namely, we choose a decomposition $\mathbb{R}^{p+q}=V_L\oplus V_R$ into a $p$-dimensional subspace $V_L$ (and a $q$-dimensional subspace  $V_R$)
together with positive quadratic forms $Q_L$ (and $Q_R$) on $V_L$ (and $V_R$) respectively, such that
we have the the left-moving and right-moving momenta $p_L$ and $p_R$:
\begin{align}
&p_L^2=Q_L(\ell)=Q(\ell)  \quad (\ell\in V_L)\;, \quad\quad
p_R^2=Q_R(\ell)=-Q(\ell) \quad (\ell\in V_R)\;.
\end{align}
We can simply write this as\footnote{Strictly speaking, $Q_L(\ell)$ was defined previously only on $V_L$, and we have now extended this to the whole of $V$ by setting $Q_L=0$ on $V_R$. A similar comment applies to $Q_R(\ell)$.}
\begin{align}
Q(\ell)=Q_L(\ell)-Q_R(\ell)=p_L^2-p_R^2 \;.
\end{align}
As in the case of the circle compactification, one can also define
a positive quadratic form,
the Hamiltonian $H(\ell):=Q_L(\ell)+Q_R(\ell)$, which can also be used as another parametrization
of the moduli space. Note that the positive quadratic form $H$ satisfies $H(\ell)\ge Q(\ell)$ for all $\ell$, and is moreover the minimal such choice;\footnote{We can define an ordering among the positive definite quadratic forms by defining $Q_1\leq Q_2$ if and only if $Q_1(\ell)\leq Q_2(\ell)$ for all $\ell$. A minimal majorant is minimal with respect to this partial ordering.}
for this reason $H$ is called a \emph{minimal majorant}. 

Notice that for any quadratic form $Q$ of signature $(p,q)$, one can
apply an element of $\text{GL}(p,q; \mathbb{R})$ to
express $Q$ in an orthonormal frame $\displaystyle \ell_1, \dots, \ell_{p+q}$:
$\ds Q(\ell)=\sum_{i=1}^p \ell_i^2 -\sum_{i=p+1}^{p+q} \ell_i^2$.
This clearly leads to $V_L=\{\ell_1, \dots, \ell_p\}$, $V_R=\{\ell_{p+1}, \dots, \ell_{p+q}\}$
and their associated quadratic forms
\begin{align}
Q_L(\ell)=\sum_{i=1}^p \ell_i^2 \;, \quad
Q_R(\ell)=\sum_{i=p+1}^{p+q} \ell_i^2\;, \quad
H(\ell)=\sum_{i=1}^{p+q} \ell_i^2\;.
\end{align}
The moduli space is parameterized by transformations that preserve $Q_L-Q_R$, modulo those that fix $H$. We also must quotient by transformations that simply permute points of the lattice. This explains the double coset in \eqref{Mpq}. 

The incompatibility between a general $\mathrm{O}(p,q; \mathbb{R})$ transformation and the integrality of the
lattice $\Lambda$
means that we have $V_L \cap \Lambda=V_R \cap \Lambda=\varnothing$
at a generic point in the moduli space.
The concepts of ``left- and right-moving lattices'' therefore do not exist.
However, there are still special sub-loci of the moduli space
where $V_L \cap \Lambda$ or $V_R \cap \Lambda$ becomes non-trivial,
and this is precisely the locus where the symmetry of the CFT enhances. 
Indeed, the moduli space $\mathcal{M}_Q$ 
arises from deformations of the Wess-Zumino-Witten models
by currents in the Cartan subalgebras of the left and right current algebra symmetries \cite{Forste:2003km}.
The cosets $\mathcal{M}_Q$ are submanifolds of $\mathcal{M}_{{\rm II}_{p,p}}$
where only restricted sets of exactly marginal operators are turned on.

\subsection{CFT Partition Function}

In the majority of this section, we study the genus one CFT partition function.  (We will comment on higher genus partition functions later in Sec.~\ref{subsec.higher_genus}.)
The genus one partition function of our theory, associated with a point $m$ of the moduli space $\mathcal{M}_Q$, can be written as
\begin{align}\label{ZQ}
Z_{Q}(\tau, \taubar;m)= \frac{\vartheta_{Q}(\tau, \taubar;m)}{\eta (\tau)^p \overline{\eta}(\taubar)^q} \;,
\end{align}
where $\tau=\tau_1+ i \tau_2$ is the modulus of the torus, $\eta(\tau)$ is the Dedekind eta function,
and $\vartheta_{Q}$ is the Siegel-Narain theta function, which is defined as:
\begin{align}\label{vartheta_def}
\vartheta_Q(\tau, \taubar;m) &:=\sum_{\ell\in \Lambda}
e^{\pi i  \tau_1 Q(\ell)  - \pi   \tau_2 H(\ell)}
=\sum_{\ell\in \Lambda}
e^{\pi i  \tau Q_L(\ell) - \pi i  \taubar  Q_R(\ell) }\;.
\end{align}
We can also write this in the more familiar notation 
\begin{align}
\vartheta_Q(\tau, \taubar;m) &=\sum_{\ell\in \Lambda}
q^{p_L^2(\ell) /2}\qbar^{p_R^2(\ell)/2}
\;,
\end{align}
with $q:=\exp(2\pi i \tau), \qbar:=\exp(2\pi i \taubar)$.
Note that this function depends explicitly on the choice of
the point $m$ in the moduli space $\mathcal{M}_{Q}$.\footnote{One might be tempted to rewrite this as a factorized expression
into sums over  ``left- and right-moving lattices'' $\Lambda_L, \Lambda_R$:
\begin{align}
\vartheta_Q(\tau, \taubar;m) &\stackrel{?}{=}
\left(\sum_{\ell_L\in \Lambda_L} q^{p_L^2(\ell)/2 } \right)
\left(\sum_{\ell_R\in \Lambda_R}\qbar^{p_R^2(\ell)/2} \right)
\;.
\end{align}
However, as remarked already such left and right-moving lattices do not exist at a generic point of the moduli space,
and therefore the theta function does not factorize into holomorphic and anti-holomorphic parts.}\\
\indent We can introduce more general partition functions.
Let us denote the dual lattice of $\Lambda$ by $\Lambda^{*}$:  
\begin{align}
\Lambda^*:= \left\{ x  \,\Big|\, Q(x, \ell) \in  \mathbb{Z} \quad  (\forall \ \ell \in \Lambda) \right\} \;.
\end{align}
By definition we have $\Lambda\subset \Lambda^*$,
but $\Lambda\subsetneq \Lambda^*$  unless $\Lambda$ is self-dual.\footnote{A lattice $\Lambda$ is self-dual if $\Lambda = \Lambda^\ast$, or equivalently if the associated quadratic form $Q$ has determinant $\pm1$, i.e. $|\det Q| = 1$.}\
Let us define the discriminant group $\mathscr{D}$ by 
\begin{align}\label{D}
\mathscr{D} := \Lambda^*/ \Lambda \;.
\end{align}

The theta function $\vartheta_{Q,h }$ shifted by a point $h  \in \mathscr{D}$ is defined as:
\begin{align}\label{theta_Qh}
\begin{split}
\vartheta_{Q,h }(\tau, \taubar;m)
&:=
\sum_{\ell\in \Lambda} e^{\pi i \tau_1 Q(\ell+h )  - \pi   \tau_2 H(\ell+h )} 
=\sum_{\ell\in \Lambda} e^{\pi i  \tau Q_L(\ell+h ) - \pi i  \taubar  Q_R(\ell+h ) } \;,
\end{split}
\end{align}
with $\vartheta_{Q}(\tau, \taubar;m) = \vartheta_{Q,h =0}(\tau, \taubar;m)$.
We can define the associated partition function as
\begin{align}\label{ZQh}
Z_{Q, h }(\tau, \taubar;m):= \frac{\vartheta_{Q, h }(\tau, \taubar;m)}{\eta (\tau)^p \overline{\eta}(\taubar)^q} \;.
\end{align}
From now on we will suppress explicit dependence of non-holomorphic quantities on $\taubar$, since this dependence should be clear from context.

The modular transformations of the Siegel-Narain theta function $\vartheta_{Q,h }(\tau;m)$ are \cite[section 4, equation (37)]{Siegel_Lecture}
\begin{equation}\label{vartheta_modular}
\begin{split}
&T: \quad \vartheta_{Q,h }(\tau+1; m) = e^{\pi i Q(h , h )} \, \vartheta_{Q,h}(\tau; m) \;,\\
&S: \quad \vartheta_{Q,h }\left(-\frac{1}{\tau};  m\right) =
\frac{ e^{-i\pi\sigma/4 }}{\sqrt{|\det Q|} }  \tau^{\frac{p}{2}} \taubar^{\frac{q}{2}} \sum_{h ' \in  \mathscr{D} }
e^{-2\pi i Q(h , h ')}  \vartheta_{Q,h '}(\tau; m) \;,
\end{split}
\end{equation}
where $\sigma:=p-q$ is the signature, and $T$ and $S$ are $\PSL(2, \mathbb{Z})$ matrices
whose $\SL(2, \mathbb{Z})$ representatives we take to be (using the same symbols $T$ and $S$)
\begin{align}
T=\left(
\begin{array}{cc}
1 & 1\\
0 & 1 
\end{array}
\right)
\;, 
\quad 
S= \left(
\begin{array}{cc}
0 & -1\\
1 & 0 
\end{array}
\right)
\;.
\end{align}
We have the relation $(ST)^3=S^2=1$ in $\PSL(2, \mathbb{Z})$. Note that the modular $S$-transformation mixes the theta functions $\vartheta_{Q,h }(\tau; m)$
with different values of $h  \in \mathscr{D}$.
Using the modular transformation rule for the eta function
\begin{equation}
\label{eta_modtrafo}
\begin{split}
T: \quad \eta(\tau+1) &= e^{2\pi i/24} \eta(\tau) \;, \\
S: \quad \eta\left(-\frac{1}{\tau} \right) &= \sqrt{-i \tau} \,  \eta(\tau) \;,
\end{split}
\end{equation}
the modular transformation of the partition function $Z_{Q, h }$ can be worked out as 
\begin{equation}\label{CFT_Z_TS}
\begin{split}
&T: \quad Z_{Q,h }(\tau+1; m) =  e^{\pi i Q(h , h )} e^{-2\pi i \sigma/24}\, Z_{Q,h }(\tau; m) \;,\\
&S: \quad Z_{Q,h }\left(-\frac{1}{\tau};  m\right) =
\frac{1}{\sqrt{|\textrm{det}\, Q|} }  \sum_{h ' \in  \mathscr{D} }
e^{-2\pi i Q(h , h ')}  Z_{Q,h '}(\tau; m) \;.
\end{split}
\end{equation}
Note that the partition function is in general not modular invariant.
This is not surprising since we are studying a general choice of the quadratic form $Q$, and in
particular the theories in general have gravitational anomalies ($p\ne q$) and also are not invariant under $S$ unless $\Lambda=\Lambda^*$. 
As we shall see, this does not affect our discussion of the ensemble average and the holographic dual.
If we impose modular invariance, 
the even lattice needs to be self-dual,
so that the discriminant group $\mathscr{D}$ is trivial.
Recall that for the self-dual case, $|\det Q|=1$. Therefore, the equations in \eqref{CFT_Z_TS} above simplify 
\begin{equation}
\begin{split}
&T:  \quad Z_{Q}(\tau+1; m) =  \, e^{-2\pi i \sigma/24}\, Z_{Q}(\tau; m) \;,\\
&S:  \quad Z_{Q}\left(-\frac{1}{\tau};  m\right) =  Z_{Q}(\tau; m) \;,
\end{split}
\end{equation}
meaning that the partition function is modular invariant if $\sigma\equiv 0$ modulo $24$.
This happens, for example, if $Q$ is one of the $24$ positive-definite, even, self-dual lattices of rank $24$, known as \emph{Niemeier lattices}. 

\subsection{Ensemble Average and the Siegel-Weil Theorem}\label{sec.SW}

Let us next consider the ensemble average of the
CFT moduli space $\mathcal{M}_{Q}$. This moduli space is a discrete quotient of a
symmetric space $G/K$ with $G=\rmO(p,q)$ and $K=\rmO(p)\times \rmO(q)$,
and has a $G$-invariant Haar measure $[dm]$,
which  is unique up to an overall multiplication by a constant.
This measure coincides with the Zamolodchikov metric of the conformal manifold (described in \eqref{coset_metric} in appendix \ref{diffeqappendix}).
Note that when integrating over the CFT moduli $m$
the moduli $\tau$ of the boundary torus will be kept to be a fixed value.

Let us consider the ensemble average of the partition function
\begin{align}\label{ensemble_Z}
\langle  Z_{Q, h }(\tau;m)  \rangle_{\mathcal{M}_Q}: = \frac{1}{\textrm{Vol}(\mathcal{M}_Q)} \,\, {\displaystyle \int_{\mathcal{M}_Q} [dm]\,\, Z_{Q, h } (\tau;m) } \;,
\end{align}
where the normalization factor 
\begin{align}
\textrm{Vol}(\mathcal{M}_Q):=\displaystyle \int_{\mathcal{M}_Q} [dm]  
\end{align}
is the volume of the moduli space.\footnote{See e.g.\ \cite{Ashok:2003gk,Moore:2015bba} for discussion of the volumes of the moduli spaces.}
Note that \eqref{ensemble_Z} is independent of the choice of the overall normalization factor of the measure on the moduli space.
Since the eta function piece of the partition function is independent of the moduli space,
this amounts to the evaluation of the ensemble average of the Siegel-Narain theta function \eqref{theta_Qh}:
\begin{align}\label{int_Qh}
\langle  \vartheta_{Q,h }(\tau)  \rangle_{\mathcal{M}_Q}: =  \frac{1}{\textrm{Vol}(\mathcal{M}_Q)} \,\,\displaystyle\int_{\mathcal{M}_Q} [d m]\,\, \vartheta_{Q,h }(\tau;m)  \;.
\end{align}
For convergence of the right hand side of \eqref{int_Qh}, we impose $p+q>4$ \cite{Siegel_Lecture}. A remarkable theorem by Siegel (see \cite[Satz 1]{MR67930} and \cite[Section 4, Theorem 12]{Siegel_Lecture}), later generalized by 
Weil \cite{weil1964certains, weil1965formule} and therefore known as the \emph{Siegel-Weil theorem}, states that when $pq\ne 0$
the ensemble average is given by
\begin{align}
\label{SiegelWeil}
\langle  \vartheta_{Q,h}(\tau)  \rangle_{\mathcal{M}_Q}
= E_{Q, h}(\tau)\;, \quad
\langle  Z_{Q, h}(\tau;m)  \rangle_{\mathcal{M}_Q} = \frac{ E_{Q, h}(\tau) }{\eta(\tau)^p \overline{\eta}(\taubar)^q} \;,
\end{align}
where  
\begin{align}\label{Siegel_Qh}
 E_{Q, h}(\tau) := \delta_{h \in \Lambda} +
 \sum_{(c,d)=1,\, c>0} \, \frac{\gamma_{Q,h}(c,d)}{ (c\tau+d)^{\frac{p}{2}} (c\taubar+d)^{\frac{q}{2}}}  
\end{align}
is the Siegel-Eisenstein series (henceforth referred to simply as Eisenstein series) associated with the quadratic form $Q$, and $\delta_{h}=1$ for $h\in \Lambda$, and $\delta_h=0$ for $h\notin \Lambda$.
Note that the constant term is expected for $h\in \Lambda$ since in the limit $\tau_2\to \infty$, we still have a contribution from the origin $\ell=0$ of the lattice $\Lambda$ in the sum \eqref{vartheta_def}.
The factor $\gamma_{Q,h}(c,d)$ is given by a version of the quadratic Gauss sum
\begin{align}\label{gamma_def}
\gamma_{Q,h}(c,d):=e^{\pi i\sigma/4} |\textrm{det} \, Q|^{-\frac{1}{2}} c^{-\frac{p+q}{2}} \sum_{\ell \in \Lambda/c\Lambda} \exp\left[ -\pi i \frac{d}{c} Q(\ell+h) \right] \;.
\end{align}
Finally, the summation is over a pair of coprime integers $c,d$ satisfying $c>0$.
Equivalently, the sum is over all rational numbers $d/c$. The modular transformations of the Eisenstein series $E_{Q,h}(\tau)$ are 
\begin{equation} 
\begin{split}
\label{modtrafo_SiegelEis}
&T:  \quad E_{Q,h}(\tau+1) =  e^{\pi i Q(h, h)} \, E_{Q,h}(\tau) \;,\\
&S:  \quad E_{Q,h}\left(-\frac{1}{\tau} \right) =
\frac{ e^{-i\pi \sigma/4 }}{\sqrt{|\textrm{det}\,Q|} }  \tau^{\frac{p}{2}} \taubar^{\frac{q}{2}} \sum_{h' \in  \mathscr{D} }
e^{-2\pi i Q(h, h')}  E_{Q,h'}(\tau) \;.
\end{split}
\end{equation}
Notice that the modular transformations of the Eisenstein series \eqref{modtrafo_SiegelEis} are equivalent to the modular transformations of the Siegel-Narain theta function described in \eqref{vartheta_modular}, as expected from the
Siegel-Weil formula \eqref{SiegelWeil}. 

We can also make contact with the results used in  \cite{Afkhami-Jeddi:2020ezh,Maloney:2020nni}. Let us consider the case of the $(p,q=p)$ Narain moduli space $\mathrm{II}_{p,p}$ discussed before.
We then have (recall again that $|\textrm{det} \, Q|=1$ for a self-dual lattice, and recall that $c$ and $d$ are coprime)
\begin{align}
\gamma_{Q_{p,p}}(c,d)=c^{-p} \sum_{n_i, w_i=0}^{c-1} \exp\left[ -2 \pi i \frac{d}{c} n_i w_i \right] =1 \;,
\end{align}
so that we have a non-holomorphic Eisenstein series
\begin{align}
\langle  \vartheta(\tau)  \rangle_{\mathcal{M}}=
E_{Q_{p,p}}(\tau):=
 \sum_{c\ge0,(c,d)=1} |c\tau+d|^{-p} \;.
\end{align}

We now give a simple proof of the Siegel-Weil theorem for an even, indefinite quadratic form $Q$. We begin by presenting the idea of the proof, which is similar to the strategy in \cite{Maloney:2020nni,Afkhami-Jeddi:2020ezh,Benjamin:2021wzr}. The first step is to show that both sides of (\ref{Siegel_Qh}) have the same behavior at the cusps of the upper half plane, which are the images of $\tau=i\infty$ under $\PSL(2,\mathbb{Z})$. Then we will derive a differential equation that is satisfied by both sides of (\ref{Siegel_Qh}), and will proceed to show that a solution to this differential equation is uniquely identified by its behavior at the cusps. The Siegel-Weil theorem then follows. This argument only relies on the transformation laws for theta functions, and does not involve explicit integration over moduli space. 

To identify the behavior of the left hand side of (\ref{Siegel_Qh}) near a cusp, note that for any $h$, the function $\vartheta_{Q,h}$ is a modular form for $\Gamma(N)$, where the level $N$ (also the level of the quadratic form $Q$) is the smallest integer such that $N Q^{-1}$ is even \cite{MR1513241,vigneras1977series}. 
The quotient $\mathbb{H}/\Gamma(N)$ has cusps at the images of $\tau=i\infty$ under $\PSL(2,\mathbb{Z})/\Gamma(N)=\PSL(2,\mathbb{Z}/N\mathbb{Z})$. The asymptotic behavior of the theta functions at $\tau=i\infty$ is given by $\delta_{h\in \Lambda}$. The asymptotic behavior of the theta functions at the other cusps is then determined by the modular transformation of the theta functions. In particular, if a modular transformation $g\in \PSL(2,\mathbb{Z})$ acts as 
\begin{align}
    \vartheta_{Q,h}(g \tau;m)=\sum_{h'\in \mathscr{D}}U_{hh'}(g,\tau)  \vartheta_{Q,h'}(\tau;m) \;,
\end{align}
then the behavior of $\vartheta_{Q,h}(\tau)$ near the cusp $\tau=g \cdot i\infty$  is given by 
\begin{align}\label{cusp_1}
\vartheta_{Q,h}\left( \tau ;m\right)\sim U_{h0}(g,g^{-1}\tau) \;.
\end{align}
The matrix $U$ can be computed from the corresponding formulas for $T$ and $S$
given previously in \eqref{vartheta_modular}, and its general formula was given in \cite{Siegel_Lecture}. If $ g\cdot\tau=(a\tau+b)/(c\tau+d)$, then we find 
\begin{align}\label{cusp_2}
\vartheta_{Q,h}\left( \tau;m \right)\sim \frac{\gamma_{Q,h}(c,-a)}{(c\tau-a)^{p/2}(c\overline{\tau}-a)^{q/2}} \;.
\end{align}
We see the behavior near each of the cusps is completely determined by the behavior at the cusp at infinity, and matches the behavior of the Eisenstein series near the cusp. 

The next step is to derive a differential equation satisfied by both sides of (\ref{Siegel_Qh}). It is simple to show that 
\begin{align}\label{diffeq}
\left(\tau_2(\partial_1^2+\partial_2^2)+\frac{p+q}{2}\partial_2+\frac{i(q-p)}{2}\partial_1\right)E_{Q,h}(\tau)=0 \;.
\end{align}
In appendix \ref{diffeqappendix}, we show that $\langle \vartheta_{Q,h}\rangle_{\mathcal{M}_Q}$ satisfies the same differential equation.
We are interested in the uniqueness of solutions to (\ref{diffeq}). For this purpose, note that if $f(\tau)$ is a solution to (\ref{diffeq}), then
\begin{align}
\left(\Box_{(p-q)/2}+\frac{((p+q)/4-1)(p+q)}{4}\right)(\tau_2^{(p+q)/4}f(\tau))=0 \;,\label{diffeq2}
\end{align}
where the weight $k$ Laplacian is defined by 
\begin{align}
\Box_k:=-\tau_2^2(\partial_1^2+\partial_2^2)+i k\tau_2\partial_1\;.
\end{align}
The minimum eigenvalue for a square normalizable eigenfunction of $\Box_k$ is\footnote{
Let $f_\lambda$ be an eigenfunction for $\Box_k$ with eigenvalue $\lambda$, assuming $k>0$ without loss of generality. Integrating by parts, we have
\begin{align}
\int_{\mathbb{H}/\Gamma(N)}\frac{d\tau_1\, d\tau_2}{\tau_2^2}\overline{f}_\lambda\left(\Box_{k}-\frac{k}{2}\left(1-\frac{k}{2}\right)\right)f_\lambda&=\int_{\mathbb{H}/\Gamma(N)}\frac{d\tau_1\, d\tau_2}{\tau_2^2}\left|\left(i\tau_2(\partial_1+i\partial_2)+\frac{k}{2}\right)f_\lambda\right|^2.
\end{align}
The right hand side is manifestly positive, so the eigenvalue of a normalizable eigenfunction of $\Box_k$ is bounded from below by
\eqref{lambda_min}.
}
\begin{align}\label{lambda_min}
\lambda_{\text{min},k}=\frac{|k|}{2}\left(1-\frac{|k|}{2}\right) \;.
\end{align}
Taking $p>q$, we then have
\begin{align}
\frac{(1-(p+q)/4)(p+q)}{4}-\lambda_{\text{min},(p-q)/2}=\frac{1}{4}(2-p)q \;.
\end{align}
We have $p>2$ for convergence, so the right hand side is less than or equal to zero. It follows that there is no normalizable eigenfunction satisfying the differential equation (\ref{diffeq2}), except in the case $q=0$ where such a function is allowed.

From now on we fix $q>0$. We consider the function
\begin{align}
f_{Q,h}(\tau)=E_{Q,h}(\tau)-\langle   \vartheta_{Q,h}(\tau)\rangle_{\mathcal{M}_Q} \;.
\end{align}
Both $E_{Q,h}$ and $\langle   \vartheta_{Q,h}\rangle_{\mathcal{M}_Q}$ are modular forms for $\Gamma(N)$ with the same eigenvalue $\lambda<\lambda_{\text{min}}$ under the Laplacian after rescaling by $\tau_2^{(p+q)/4}$. Therefore $f_{Q,h}$ is as well. But the asymptotics of $\langle   \vartheta_{Q,h}\rangle_{\mathcal{M}_Q}$ and $E_{Q,h}$ are the same at the cusps, so $f_{Q,h}$ is zero at all of the cusps. Since there cannot be a normalizable eigenfunction of the Laplacian on $\mathbb{H}/\Gamma(N)$ with eigenvalue $\lambda<\lambda_{\text{min}}$, it follows that $f_{Q,h}=0$, which completes the argument.

\subsection{Bulk Interpretation}

Having identified the ensemble average, let us now come to the holographic interpretation.
In the holographic bulk we expect a sum over semiclassical geometries which are asymptotically AdS$_3$.
Such geometries were classified in \cite{Maloney:2007ud},\footnote{This amounts to the classification of hyperbolic 3-manifolds 
$\mathbb{H}^3/\Gamma$ whose fundamental group is contained in that of the two-dimensional boundary torus.}
and include geometries labeled by an element of $\PSL(2,\mathbb{Z})$,
the so-called ``$\PSL(2, \mathbb{Z})$ black holes'' \cite{Maldacena:1998bw,Dijkgraaf:2000fq}.\footnote{The asymptotically AdS$_3$ boundary condition allows for orbifolds $M_{(c,d)}/\mathbb{Z}_m$ of $\PSL(2, \mathbb{Z})$ black holes  \cite{Maloney:2007ud}. There is in general no consensus on which geometries we should include in the path integral of fully quantum gravity. We will not include these orbifold geometries in this paper, since these geometries are not needed for reproducing our partition functions.}
Mathematically, these are solid tori with torus boundaries (genus one handlebodies),
where $\PSL(2, \mathbb{Z})$ acts as the mapping class group
on the boundary torus. More precisely,
the geometry is labeled by $\Gamma_{\infty}\backslash \PSL(2, \mathbb{Z})$,
where $\Gamma_{\infty}\simeq \mathbb{Z}$
is the Abelian group generated by the matrix $T$. An element of the quotient group $\Gamma_{\infty}\backslash \PSL(2, \mathbb{Z})$
can be parametrized by a pair of coprime integers $(c,d)$ with $c>0$, since given such a pair we can uniquely identify an element $\left(\begin{array}{cc} a & b \\c & d\end{array} \right)$ of $\PSL(2, \mathbb{Z})$, up to ambiguities in 
$\Gamma_{\infty}$. We denote the associated geometry by $M_{(c,d)}$:
these geometries include thermal AdS$_3$ ($M_{(0,1)}$) and the BTZ black hole ($M_{(1,0)}$) \cite{Banados:1992wn}.
Since we have a sum over essentially the same pair $(c,d)$ in \eqref{Siegel_Qh},
we expect to interpret the sum in \eqref{Siegel_Qh} as a sum over geometries.

One subtlety for us is that  that our theories in general have gravitational anomalies (since $p\ne q$), 
and hence the partition function is not invariant under the large coordinate transformations in $\Gamma_{\infty}$.
(Relatedly, the BTZ black hole is now rotating with angular momentum $\displaystyle J = \frac{\sigma}{24} = \frac{p-q}{24}$ \cite{Kraus:2006wn}.) 
In the discussion of the partition function, we need to be careful in picking up a representative from the coset
$\Gamma_{\infty}\backslash \PSL(2, \mathbb{Z})$, since different choices give partition functions 
differing by factors of $\exp(2\pi i \sigma/24)$. 

By identifying the $\delta_{h\in \Lambda}$ piece 
as a contribution from matrices with $c=0$ (and hence $d= 1$),
we can write
\begin{align}
\langle Z_{Q,h}(\tau) \rangle_{\mathcal{M}_Q}
&=  \frac{1}{\eta( \tau)^p \bar\eta(\taubar)^q}  \sum_{(c,d)=1,c\ge 0} \frac{\gamma_{Q,h}(c,d)}{(c\tau+d)^{\frac{p}{2}}(c\taubar+d)^{\frac{q}{2}}} \;,
\end{align}
where we defined $\gamma_{Q,h}(0, 1):=\delta_{h\in \Lambda}$.
Owing to the modular transformations of the Dedekind eta function mentioned previously in \eqref{eta_modtrafo}, one obtains 
\begin{align}
\langle Z_{Q,h}(\tau) \rangle_{\mathcal{M}_Q}&= \sum_{g\in \Gamma_{\infty}\backslash \PSL(2, \mathbb{Z})}
e^{\frac{2\pi i \sigma}{24} \Phi(g)-\frac{i\pi \sigma}{4}}
\frac{\gamma_{Q,h}(c,d)}{\eta(g \cdot \tau)^p \bar \eta(g\cdot \taubar)^q}  \;,
\label{gamma_sum}
\end{align}
where $g$ is a $\PSL(2, \mathbb{Z})$ matrix of the form $\left(\begin{array}{cc} a & b \\c& d \end{array}\right)$,
and $\Phi(g)\in \mathbb{Z}$ is the Rademacher function.\footnote{
The modular transformation rule for the eta function is given by
\begin{align}\label{eta_modular}
\eta\left(
\frac{a\tau+b} {c\tau+d}
\right)
&=\exp\left(\frac{2\pi i}{24} \Phi \left(
\begin{array}{cc}
a & b \\
c & d
\end{array}
\right) \right) (-i(c\tau+d))^{\frac{1}{2}} \eta(\tau) \hspace{10 mm} c>0 \;.
\end{align}
Here the Rademacher function $\Phi(g)\in \mathbb{Z}$ is defined by
\begin{align}
\Phi \left(
\begin{array}{cc}
a & b \\
c & d
\end{array}
\right)
=
\frac{a+d}{c}-12  s(d, c) \hspace{10 mm} c>0
 \;,
\end{align}
and the Dedekind sum $s(d,c)$ for $c>0$ is defined by
\begin{align}
s(0,1):=0  \;, \qquad
s(d,c):=\sum_{k=1}^{c-1} \left(\!\!\left( \frac{k}{c} \right) \!\!\right)\left(\!\!\left( \frac{dk}{c} \right) \!\!\right)\;,
\end{align}
with
\begin{align}
\left(\!\left( x \right) \!\right)
:=\begin{cases}
0 & (x\in \mathbb{Z}) \\
x-[x]-\frac{1}{2} & (\textrm{otherwise})
\end{cases}
\;.
\end{align}
}
Note that the phase factor $\exp(2\pi i \sigma \Phi(g)/24)$
as well as the eta functions $\eta(g \cdot \tau)^p \bar \eta(g\cdot \taubar)^q$
depend on the choice of a representative of the quotient $\Gamma_{\infty}\backslash \PSL(2, \mathbb{Z})$, as expected from the gravitational anomaly. However, the whole combination does not depend on such a choice.

Let us next consider the contribution from the thermal AdS$_3$ geometry.
While the graviton has no dynamical degrees of freedom in the bulk of
the three-dimensional gravity, there are
boundary excitations, as
studied by Brown and Henneaux \cite{Brown:1986nw}.
In our context, we can construct boundary 
Virasoro generators by the Sugawara construction \cite{Sugawara:1967rw} of the $\U(1)^{p+q}$ current algebras,
$p$ left- and $q$ right-moving. We therefore expect the partition function to be
\begin{align}
Z[M_{(1,0)}] \stackrel{?}{=} \frac{1}{\eta(\tau)^p \bar\eta(\taubar)^q} \;,
\end{align}
and by summing over $\PSL(2, \mathbb{Z})$ images
we might expect
\begin{align}\label{Z_bulk?}
Z_{\rm bulk} \stackrel{?}{=} \sum_{g \in \Gamma_{\infty}\backslash \PSL(2, \mathbb{Z})} \frac{1}{\eta(g\cdot \tau)^p \overline{\eta}(g\cdot \taubar)^q}  \;.
\end{align}
In fact, this is precisely the logic which worked for the 
special case of the ${\rm II}_{p,p}$ lattice \cite{Maloney:2020nni}.
In this special case, we have
\begin{align}
\langle  Z_{{\rm II}_{p,p}}(\tau)  \rangle_{\mathcal{M}}=
\frac{E_{{\rm II}_{p,p}}(\tau)}{|\eta( \tau)|^{2p} }  =
 \sum_{(c,d)=1,c\ge 0}  \frac{1}{| c\tau+d|^p  |\eta( \tau)|^{2p} }\;.
\end{align}
and the expression \eqref{gamma_sum} has no ambiguities:
\begin{align}\label{theta_II_p}
\langle  \vartheta_{{\rm II}_{p,p}}(\tau)  \rangle_{\mathcal{M}}=
 \sum_{g \in \Gamma_{\infty}\backslash \PSL(2, \mathbb{Z})}  \frac{1}{  |\eta(g \cdot \tau)|^{2p} }\;,
\end{align}
as anticipated in \eqref{Z_bulk?}. Moreover, the contribution from each geometry
was identified with the partition function of the three-dimensional Abelian Chern-Simons theory,
whose gauge group is $\U(1)^{p+q}$ and whose Lagrangian (in Euclidean signature) is determined by the quadratic form $Q$
\begin{align} 
S_{\rm CS}\label{CS}
&=\sum_{i,j=1}^{p+q} \frac{i}{8\pi} Q_{i,j } \int_M A^i \wedge d A^j 
= \frac{i}{8\pi} \int_M  Q(A, dA)  \;.
\end{align}
(Recall that $Q$ is even, ensuring the integer quantization of the levels.)
Note that the appearance of the $\U(1)^{p+q}$ gauge symmetry in the bulk is expected
from the $\U(1)^{p+q}$ global symmetry of the boundary theory, and the existence of the Chern-Simons term
is suggested from the anomalies of the global symmetries. Moreover the eta function contributions in \eqref{theta_II_p}
were derived from the one-loop analysis of the Chern-Simons theory, building on similar computations 
in three-dimensional gravity \cite{Giombi:2008vd}.\footnote{There are, however, potential subtleties associated with the asymptotic boundary conditions of the fields in the Chern-Simons theory.} 

Our discussion for a general, even quadratic form $Q$ is more involved
than the special case of the ${\rm II}_{p,p}$ lattice, as is evident, e.g.,\ from the non-trivial factors $\gamma_{Q,h }(c,d)$
in \eqref{Siegel_Qh}. It turns out, however,
that the bulk theory is still described by the Abelian Chern-Simons theory \eqref{CS}
in our more general setting. While the bulk theory is an exotic theory of gravity whose complete understanding is beyond the scope of this paper, the Abelian Chern-Simons theory is a good approximation to the exotic theory and 
will successfully reproduce many results, including the phase factor $\gamma_{Q,h }(c,d)$.

Incidentally, in the condensed matter literature these Chern-Simons theories are used for the classification of 
topological phases of interacting system in two spatial dimensions \cite{Lu:2012dt},
where the matrix $Q$ is often called the $K$-matrix.\footnote{In the literature the level of the Chern-Simons theory is 
often denoted by $Q/2$, not $Q$. Our normalization here is useful when we discuss spin Chern-Simons theory in section \ref{sec.fermionic}.\label{footnote.CS_normalization}}
It is remarkable that all such theories arise from ensemble averages discussed in this paper.

In order to derive the phase factor $\gamma_{Q,h }(c,d)$, 
let us first recall the canonical quantization of the $\U(1)$ Chern-Simons theory with integer level\footnote{As in footnote \ref{footnote.CS_normalization}, we choose a normalization 
where the minimal value of the level for the non-spin (even) Chern-Simons theory is $k=2$.} $k$ is spanned by a set of states $|h  \rangle$ ($h=0, 1/k, \dots, (k-1)/k$) corresponding to 
a path-integral over a solid torus with an insertion of a charge $kh$ Wilson line inside.
The modular group is represented on the Hilbert space by the operators
\begin{align}
\begin{split}
&\bm{T}  | h \rangle =
 e^{\pi i kh^2} e^{-2\pi i /24}\, | h \rangle \;,\\
&\bm{S} | h \rangle=
\frac{1}{\sqrt{k} }  \sum_{h' \in  \mathscr{D} }
e^{-2\pi ikh h'}  | h'   \rangle\;.
\end{split}
\end{align}
Note that the phase factor in the action of $T$ represents the framing anomaly of the Chern-Simons theory \cite{Witten:1988hf},
while that of $S$ is simply a discrete Fourier transformation.
It is straightforward to work out a similar formula for a more general Abelian Chern-Simons theory \eqref{CS},
so that one has 
\begin{align}\label{CS_TS}
\begin{split}
&\bm{T}  | h; m \rangle = \, e^{\pi i Q(h, h)}
e^{-2\pi i \frac{\sigma}{24}}\, |h ; m \rangle \;,\\
&\bm{S} | h;  m \rangle=
\frac{1}{\sqrt{|\det Q|} }  \sum_{h' \in  \mathscr{D} }
e^{-2\pi i Q(h, h')}  | h' ; m  \rangle\;,
\end{split}
\end{align}
where $h, h'$ are elements of the discriminant group $\mathscr{D}$ \eqref{D}.
Since $\SL(2, \mathbb{Z})$ is generated by $S$ and $T$-transformations,
one can work out the action of a more general matrix $g=\left(\begin{array}{cc} a& b\\ c&d \end{array} \right)\in \SL(2,\mathbb{Z})$:
\begin{align}
\bm{U}(g)  | h\rangle = \sum_{h' \in  \mathscr{D} }
 (\bm{U}(g))_{h, h'} |h'  \rangle \;.
\end{align}

Now coming back to the discussion of holography, 
we are interested in the geometry of the solid torus without any Wilson line insertions,
namely in the state $|h=0\rangle$. 
This means that we are interested in the matrix element $(\bm{U}(g))_{0,h}$,
which we find to be related by complex conjugation to the factor $\gamma_{Q, h}(c,d)$ in the Eisenstein series $E_{Q,h}$
(see also the discussion around \eqref{cusp_1} and \eqref{cusp_2}):
\begin{align}
\langle 0|\bm{U}(g)|h\rangle^*=\langle h|\bm{U}(g)^{-1}|0\rangle =e^{\frac{2\pi i\sigma \Phi(g)}{24}-\frac{i\pi\sigma}{4}} \gamma_{Q, h}(c,d)\;,\hspace{ 10 mm}c>0 \;.\label{ugammamatch}
\end{align}
As this discussion makes clear, the expression $(\bm{U}(g))_{h,0}$
in itself can be identified as the Chern-Simons partition function
of the geometry obtained by gluing two solid tori along the boundary torus
by the mapping class group element represented by $\bm{U}(g)$.
This is the lens space $L(c,d)$,
which is defined for $c\ne 0$ by a discrete $\mathbb{Z}_c$ quotient of the three-sphere $|z_1|^2+|z_2|^2=1$ (with complex $z_1, z_2$) by
\begin{align}
(z_1, z_2) \sim (e^{2\pi i \frac{1}{c}} z_1, e^{2\pi i \frac{d}{c}} z_2) \;;
\end{align}
for $c=0$ the lens space is defined to be $S^1\times S^2$.
The Chern-Simons partition function for lens spaces was computed by Jeffrey in \cite{MR1175494}, and also in various other papers such as \cite{lawrence1999witten,hansen2004reshetikhin,Beasley:2005vf,Blau:2006gh,jeffrey2011nonabelian,Kallen:2011ny,Guadagnini:2014mja,Gang:2019juz}.
The parameters $h, h'$ represent insertions of 
Wilson lines in each solid torus.
The extra phase factor $\exp(2\pi i\sigma \Phi(g)/24)$
represents the effect of the framing anomaly.
The partition functions of the lens spaces, without any Wilson line insertions, are given by $(\bm{U}(g))_{0, 0}$,
which is expressed as a sum over contributions from flat connections---for each contribution,
the phase represents the $\eta$-invariant of the three-manifold \cite{MR397797},
or equivalently the phase in the one-loop determinant for the Chern-Simons theory \cite{Witten:1988hf}.
In summary, 
\begin{align}
\label{eisensteininvariants}\langle Z_{Q,h}(\tau) \rangle_{\mathcal{M}_Q}
=\sum_{g\in \Gamma_{\infty}\backslash \PSL(2, \mathbb{Z})}
\frac{\langle h | \bm{U}(g)^{-1} | 0\rangle}{\eta(g \cdot \tau)^p \overline{\eta}(g\cdot \taubar)^q}  \;.
\end{align}
This completes our derivation of the bulk partition function.

In our discussion of the matrix elements of $\bm{U}(g)$, it was crucial to assume that the gauge group of the Chern-Simons theory 
is $\U(1)^{p+q}$, not $\mathbb{R}^{p+q}$. If we wish to obtain the honest wavefunction
of the $\U(1)^{p+q}$ Chern-Simons theory, however, we should rather consider a sum of the expression \eqref{eisensteininvariants}
over large gauge transformations of the gauge fields. We will then have a theta function in the numerator, to match with the character of the 
boundary current algebra. (We will discuss such wavefunctions in a more generalized setup in section \ref{sec.before}.)
The choice of the bulk gauge group, $\U(1)^{p+q}$ or $\mathbb{R}^{p+q}$, is therefore a subtle question (see \cite{Maloney:2020nni} for related discussion), and 
we will leave a better concentual understanding of this subtlety for future works.

It is interesting to notice that quadratic forms with different signatures $(p, q)$ are related by 
analytic continuation. In other words, CFT moduli spaces with different signatures are
all included when we 
analytically continue the gauge group \cite{Witten:2010cx}
$\U(1)^{p+q}$ to $(\mathbb{C}^{\times})^{p+q}$
in the Chern-Simons theory; different choices of the signatures arise by choosing different integration contours.

\subsection{Positive Definite Case}

In the discussion of the Siegel-Weil formula the special case of $pq=0$ was excluded
when we stated the formula. In the chiral case ($q=0$), the moduli space $\mathcal{M}_Q$ is zero-dimensional
and therefore trivial. It turns out that there is still a formula of the form \cite{MR1503238}
\begin{align} \label{Siegel_p=0}
\langle\! \langle  \vartheta_Q(\tau) \rangle\! \rangle =E_{Q}(\tau) \;.
\end{align}
However the ensemble average here, represented by the symbol $\langle\! \langle - \rangle\! \rangle$,
is different from those for the cases $pq\ne 0$ - instead of fixing a quadratic form
we have a sum over different quadratic forms in the ``class'' of $Q$.
To explain this we introduce some terminology. Two even quadratic forms $Q,Q'$ are equivalent in a field $\mathbb{F}$
 if there exists an element $g$ of $\GL(p+q;\mathbb{F})$
such that $Q'= g^{T} Q g$. 
We say that $Q, Q'$ are in the same class if the two quadratic forms are 
equivalent in $\mathbb{Z}$.
Similarly, $Q, Q'$ are in the same genus if $Q$ and $Q'$ are equivalent in
$\mathbb{R}$ as well as $\mathbb{Z}_p$ for all prime $p$. 
It is known that $Q$ and $Q'$ are in the same genus if and only if
we have $Q\oplus \mathrm{II}_{1,1} \simeq Q'\oplus \mathrm{II}_{1,1}$.
There are only a finite number of classes inside a given genus $g(Q)$,
and this is the class number $h(Q)$.

For a given $Q$ we can consider a representative class $Q_1, \dots, Q_{h(Q)}$ with the same genus
as $Q$. Since the Siegel-Narain theta function depends only on the class of $Q$,
the set of theta functions $\{ \vartheta_{Q_j} \}$ do not depend on the
choice of representative elements from the genus of $Q$.

The ensemble average in \eqref{Siegel_p=0} is defined by a weighted sum
\begin{align}
\langle\! \langle \vartheta_Q(Z) \rangle\! \rangle:=\frac{1}{M(Q)}\displaystyle\sum_{j=1}^{h(Q)} \frac{\vartheta_{Q_j}(Z) } {|\rmO_{Q_j}(\mathbb{Z})|} 
 \;, \quad M(Q) := \displaystyle\sum_{j=1}^{h(Q)} \frac{1 } {|\rmO_{Q_j}(\mathbb{Z})|} \;,
\end{align}
where the normalization factor $M(Q)$ is known as the mass of the quadratic form $Q$.
Note that holography for chiral theories was discussed in \cite{Dymarsky:2020bps},
see also \cite{Dymarsky:2020pzc}. It is far from clear physically, however, why we need to consider such an ensemble average.\footnote{For indefinite cases there is only one class in a given genus \cite{MR1245266}.} As an example, for the case $p=24$ this ensemble average is a sum over the $24$ even self-dual lattices, the Niemeier lattices.

Since there are no continuous moduli for a positive definite lattice, we can consider observables such as correlation functions in addition to the partition function. Here we will give one example, fixing the self-dual case for simplicity. Let $P(\ell)$ be a polynomial which is spherical with respect to $Q$, meaning that $Q^{ij}\partial_i \partial_j P=0$. Then $P(\partial X)$ is a primary operator in the conformal field theory, and we can consider its one-point function on the torus \cite{MR1827085}. This correlation function is equal to a spherical theta function, 
\begin{align}
\vartheta_{Q,P}(\tau)=\sum_{\ell\in \Lambda}P(\ell)e^{iQ(\ell)\tau} \;.
\end{align}
A theorem of Waldspurger \cite{eichlerzagier,waldspurger} computes the ensemble average of $\vartheta_{Q,P}$ for some specific spherical polynomials,
\begin{align}
\langle\! \langle \vartheta_{Q,P^\nu_m}\rangle\! \rangle=C_k^{(\nu)}|T_m \;.
\end{align}
Here $P^\nu_m$ is defined in terms of Gegenbauer polynomials, $T_m$ is the Hecke operator, and $C_k^{(\nu)}$ is known as Cohen's function (see \cite{eichlerzagier} for details). It would be interesting to understand the bulk interpretation of these correlation functions.

\subsection{Higher Genus} \label{subsec.higher_genus}

We can repeat the discussions above for a higher genus boundary surface $\Sigma_g$. The higher genus theta function is given by the expression 
\begin{align}
\vartheta^g_{Q,\vec{h}}(\Omega;m) &:=\sum_{\vec{\ell}\in \Lambda^g}
e^{\pi i  \text{Tr}(\Omega_1 Q(\vec{\ell}+\vec{h}))  - \pi   \text{Tr}(\Omega_2 H(\vec{\ell}+\vec{h}))}
=\sum_{\vec{\ell}\in \Lambda^g}
e^{\pi i \text{Tr}( \Omega Q_L(\vec{\ell}+\vec{h})) - \pi i  \text{Tr}(\bar{\Omega}  Q_R(\vec{\ell}+\vec{h}))}\;,
\end{align}
where $\Omega=\Omega_1+i \Omega_2$
is the period matrix of size $g\times g$ that parametrizes the Siegel upper half plane, and $\vec{\ell}=(\ell_1, \dots, \ell_g)$.
The averaged partition function, derived in \cite{MR10574}, is given by 
\begin{align}\label{SW_higher}
\langle \vartheta^g_{Q,\vec{h}}(\Omega;m)\rangle_{\mathcal{M}_Q}&=E^g_{Q,h}(\Omega)\;,
\end{align}
where 
\begin{align}
E^g_{Q,\vec{h}}(\Omega)=\sum_{\gamma\in \Gamma_\infty \backslash \Sp(2g,\mathbb{Z})}\frac{\gamma_{\vec{h}}(C,D)}{\det(C\Omega+D)^{p/2}\det(C\overline{\Omega}+D)^{q/2}} \;
\end{align}
is the Siegel-Eisenstein series and we have assumed that $p+q>2g+2$. 
The expressions for $\gamma_{\vec{h}}(C,D)$, which generalizes the genus $1$ expressions in \eqref{gamma_def}, can be
found in \cite[Section 12]{MR10574}. The summation in \eqref{SW_higher}
is equivalently over the Lagrangian sublattices in $H^1(\Sigma_g, \mathbb{Z}_2)$ \cite{Maloney:2020nni}, 
and when $\Sigma_g$ is connected can be identified with a sum over handlebodies.
We then expect that $\gamma_{\vec{h}}(C,D)$ should be matched with the partition functions of 
the Abelian Chern-Simons theory \eqref{CS}, now on 3-manifolds obtained by gluing two genus $g$ handlebodies (i.e.\ 3-manifolds with Heegaard genus $g$).\footnote{In three-dimensional gravity there are classical solutions
with conformal boundary which are not handlebodies \cite{Yin:2007at}. It seems that these geometries do not contribute to
the partition function. The identification of the bulk geometry is more non-trivial when the boundary geometry $\Sigma$ has several disconnected components, see \cite{Maloney:2020nni} for further discussion.}
Note that for any choice of $p$ and $q$ the formula \eqref{SW_higher} holds only for finitely many $g$'s---since the exotic bulk theory is ``coarse-grained,'' it is not surprising that we have access to only finitely many invariants. 

\section{Ensemble Average of Fermionic CFTs}\label{sec.fermionic}

In this section we extend the discussion of the previous section 
to an integral quadratic form $Q$ which is not necessarily even.
While this might look like a minor change, 
such a generalization requires us to carefully discuss
spin-structure dependence of 
our holographic  dualities.\footnote{See \cite{Balasubramanian:2020jhl} for a recent discussion of spin structures in 
two-dimensional quantum gravity.}

\subsection{Review of Spin Chern-Simons Theory}

To explain the spin-structure dependence, let us begin with the bulk 
Chern-Simons theory. 

Let us recall the standard argument for the quantization of the 
levels of the Chern-Simons theory.  
While the Chern-Simons Lagrangian \eqref{CS} is apparently not gauge-invariant,
one can consider a four-manifold $N$ bounding the three-manifold $M$,
to rewrite the action \eqref{CS} in terms of the gauge-invariant field strength $F=dA$ as 
\begin{align} 
S_{\rm CS}\label{CS_Poincare}
= \frac{i}{8\pi} \int_{M=\partial N}  Q(A, F) 
= 2\pi \int_{N}  \frac{i}{8\pi^2} \frac{Q(F, F) }{2}  \;.
\end{align}
While this depends on the choice of $N$, a different choice $N'$
gives an answer which differs from that of $N$ by 
\begin{align} 
\Delta S_{\rm CS}\label{CS_Delta}
= 2\pi i \int_{N'\cup \bar{N}}  \frac{1}{8\pi^2} \frac{Q(F, F) }{2}  \;,
\end{align}
where $N'\cup \bar{N}$ is a closed four-manifold obtained by gluing $N'$ and $\bar{N}$ ($N$ with 
orientation reversed) along the common three-manifold $M$.
Since $Q$ is even, $Q/2$ is integral and $ Q(F, F)/(2\times 8\pi^2)$
gives an element of the integer cohomology class. 
This means that the combination $\exp(- S_{\rm CS})$ is gauge-invariant in the path integral.

This argument does not apply for odd integral $Q$. We can, however, cure this problem by 
requiring that the three manifold $M$ is spin, and by requiring that the 
bounding four-manifold $N$ admits a spin structure compatible with that of $M$ \cite{Belov:2005ze,dijkgraaf}.
Then the integral of 
$\displaystyle \frac{Q(F, F)}{2}\frac{1}{8\pi^2}$ over a closed spin four-manifold is now an integer for any integral quadratic form $Q$.
The resulting theory
then depends on both the topology of the three-manifold $M$, as well 
as a choice of the spin structure on it.\footnote{Any compact oriented three-manifold admits a spin structure \cite{MR1809834}.}
We call this theory the spin Chern-Simons theory (associated with an integral quadratic form $Q$). 
Notice that while we are expected to sum over all the possible geometries in the theories of quantum gravity,
one can still restrict the geometries by fixing their spin structures.

In the rest of this section, we will discuss how to incorporate this spin-structure dependence into the framework of ensemble averages.

\subsection{Partition Functions with Spin Structure}

In the boundary theory, we consider CFTs that are dependent on the choice of the spin structure,
namely fermionic CFTs (spin CFTs).\footnote{See \cite{Kapustin:2017jrc,Tachikawa_Lec,Karch:2019lnn} for recent discussion on spin CFTs.}

\begin{figure}[htbp]
	\centering
	\scalebox{0.9}{
		\begin{tikzpicture}[every node/.style={}]
		\node (a11) at (0,3)  {$(1,1)$};
		\node (a01) at (3,3)  {$(0,1)$};
		\node (a10) at (3,0)  {$(1,0)$};
		\node (a00) at (0,0)  {$(0,0)$};
		\path[<-] 
		(a11) edge [in=90, out=150, loop, above left] node {$T$} ()
		(a11) edge [in=300, out=0, loop, above right] node {$S$} ()
		(a01) edge [in=130, out=50, loop, above] node {$T$} ()
		(a00) edge [in=150, out=210, loop, left] node {$S$} ()
		(a00) edge [bend left] node [below] {$T$} (a10) 
		(a10) edge [bend left] node [below] {$T$} (a00) 
		(a01) edge [bend right] node [left] {$S$} (a10) 
		(a10) edge [bend right] node [right] {$S$} (a01) 
		;
		\end{tikzpicture}
	}
	\caption{Change of spin structures under the mapping class group transformations $T, S$. 
	Three even spin structures $(0,0), (0,1), (1,0)$ make a triplet, while the odd spin structure $(1,1)$ is a singlet.}
	\label{map}
\end{figure}
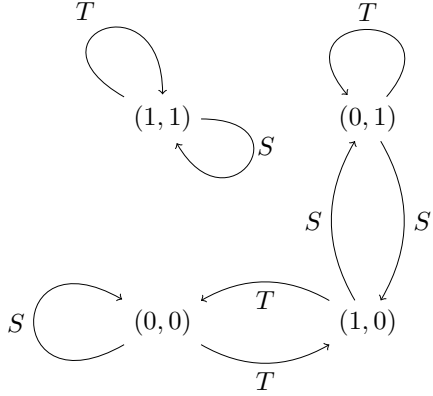

There are four spin structures on the boundary two-dimensional torus,
which are labeled by $H^1(\mathbb{T}^2; \mathbb{Z}_2)=\mathbb{Z}_2\oplus \mathbb{Z}_2$.
We will denote this by the $\mathbb{Z}_2$-signs $\epsilon_1, \epsilon_2$, each of which takes values in $0$ and $1$. 
Following \cite{Belov:2005ze} we define the
generalization of the theta function \eqref{theta_Qh} to be
\begin{align}\label{vartheta_spin}
\vartheta^{\epsilon_1,\epsilon_2}_{Q,h}(\tau,m):=\sum_{\ell\in \Lambda+h+\epsilon_1 W/2}e^{i\pi \tau Q_L(\ell)-i\pi \overline{\tau}Q_R(\ell)}(-1)^{\epsilon_2(W,\ell)} \;,
\end{align} 
where $h$ is an element of $\mathscr{D}=\Lambda^*/\Lambda$, just as before. The characteristic class $W\in \Lambda^*$, 
known as the (integral) Wu class \cite{MR0440554,MR0145525}, is defined by $(W,\ell)\equiv Q(\ell)\text{ mod }2$ for $\ell\in \Lambda$.\footnote{This condition determines only the 
element of the quotient $[W] \in \Lambda^*/(2\Lambda^*)$.}
This is solved by $W_\alpha=Q_{\alpha\alpha}$.\footnote{We can check this for binary forms, for example. If $Q(\ell)=a\ell_1^2+2b\ell_1\ell_2+c\ell_2^2$, then $W_\alpha=(a,c)$, so $(W,\ell)=a\ell_1+c\ell_2$. Reducing mod 2 we see that $Q(\ell)\equiv (W,\ell)$.} Note that only $\vartheta^{0,0}_{Q,h}$ and $\vartheta^{0,1}_{Q,h}$ are non-vanishing at the cusp at $\tau=i\infty$.
The modular transformations of the theta functions are given in appendix \ref{sec.oddtheta}.
Note that the modular transformations mix spin structures, as shown in Fig.~\ref{map}.

The theta function above \eqref{vartheta_spin} (with $h=0$) is reminiscent of the free fermion partition function
\begin{align}
\theta^{\epsilon_1,\epsilon_2}(\tau)=\sum_{\ell\in \mathbb{Z}+\epsilon_1/2}q^{\ell^2}(-1)^{\epsilon_2\ell^2}.
\end{align}
Here $\epsilon_1$ labels the periodicities on the spatial circle (R or NS), and $\epsilon_2$ labels the periodicity on the thermal circle.
In the Narain CFT, we have operators $:\exp(ik_L\cdot \hat{X}_L+ik_R\cdot \hat{X}_R):$. These are fermions if $Q(k)$ is odd and are bosons if $Q(k)$ is even \cite{Polchinski:1998rr}. Now let us show that \eqref{vartheta_spin} is the partition function of the CFT. The dependence on $\epsilon_2$ is obvious. To derive the dependence on $\epsilon_1$, recall the mode expansion for a boson,
\begin{align}
\hat{X}_L&=-i\hat{p}_L\log z+\text{analytic}\;,\\
\hat{X}_R&=-i\hat{p}_R\log \overline{z}+\text{analytic}\:.
\end{align}
Therefore as we go around the circle, we have 
\begin{align}
(\hat{X}_L,\hat{X}_R)\to(\hat{X}_L+2\pi \hat{p}_L,\hat{X}_R-2\pi \hat{p}_R)\;.
\end{align}
We want to find the analog of a spin field for a free fermion, which makes fermions anti-periodic on the spatial circle \cite{Polchinski:1998rr}. We make an ansatz of exponential form $:\exp(ip\cdot \hat{X}):$. As we go once around the spatial circle, this transforms as
\begin{align}
:e^{ip\cdot \hat{X}}:(e^{2\pi i }z)=e^{2\pi i Q(p,\hat{p})}:e^{ip\cdot \hat{X}}:(z) \;.
\end{align}
Now consider the state $:\exp(ip\cdot \hat{X}):(z)|\ell\rangle$, where $\ell\in \Lambda$.  When we go around the circle, this state picks up a phase $\exp(2\pi iQ(p,\ell))$. In order for this state to be anti-periodic when $\ell$ is a fermion and periodic when $\ell$ is a boson, we therefore need $2Q(p,\ell)\equiv Q(\ell)\text{ mod }2$. But this is precisely the definition of the characteristic class $W$. Therefore the spin field is $:\exp(i W \cdot \hat{X}/2):$. This creates the Ramond sector ground state, and the dependence on $\epsilon_1$ then follows.

\subsection{Ensemble Average}
Now that we have computed the partition function for a spin CFT, we can define the Eisenstein series with spin structure $(\epsilon_1,\epsilon_2)$ as the average of $\vartheta_{Q,h}^{\epsilon_1,\epsilon_2}$ over the moduli space, 
\begin{align}
E^{\epsilon_1,\epsilon_2}_{Q,h}(\tau):=\langle \vartheta^{\epsilon_1,\epsilon_2}_{Q,h}(\tau)\rangle_{\mathcal{M}_Q} \;.
\end{align}
As in the even case, we would like to understand the interpretation of this Eisenstein series in terms of the bulk Chern-Simons theory.

Let us start with the case of $\epsilon_1=\epsilon_2=0$.
In this case, the definition \eqref{vartheta_spin} coincides with the previous definition in the even case \eqref{vartheta_def},
except here $Q$ is not necessarily even. We can cure this problem by writing \eqref{vartheta_spin}
in terms of $2Q$ and $\tau/2$:
\begin{align}
\vartheta^{0,0}_{Q,h}(\tau,m)
=\sum_{\ell\in \Lambda+h}e^{i\pi \frac{\tau}{2} (2Q_L(\ell))-i\pi \frac{\overline{\tau}}{2} (2Q_R(\ell))}=
\vartheta_{2Q,h}\left(\frac{\tau}{2},m\right)\;.
\end{align} 
Since this is expressed in terms of the theta function for an even quadratic form, the ensemble average can be evaluated using the analysis of section \ref{sec.SW}. We find
\begin{align}
E^{0,0}_{Q,h}(\tau,m) 
=\langle \vartheta^{0,0}_{Q,h}(\tau,m) \rangle_{\mathcal{M}_Q}
=\left\langle \vartheta_{2Q,h}\left(\frac{\tau}{2},m\right) \right\rangle_{\mathcal{M}_{2Q}}
=E_{2Q,h}\left(\frac{\tau}{2},m\right) \;.
\end{align} 
In terms of the sum over geometries, we have
\begin{align}
E_{Q,h}^{0,0}(\tau)&=\delta_{h\in \Lambda}+\sum_{(c,d)=1,c>0}\frac{2^{(p+q)/2} \gamma_{2Q,h}(c,d)}{(c\tau+2d)^{p/2}(c\overline{\tau}+2d)^{q/2}}\;.
\end{align}
Since $2^{(p+q)/2} \gamma_{2Q,h}(c,d) = \gamma_{Q,h}(c,2d)$, from the definition of $\gamma_{Q,h}(c,d)$ in \eqref{gamma_def},
we obtain
\begin{align}
E_{Q,h}^{0,0}(\tau)&=\delta_{h\in \Lambda}+\sum_{(c,d)=1,c>0}\frac{\gamma_{Q,h}(c,2d)}{(c\tau+2d)^{p/2}(c\overline{\tau}+2d)^{q/2}} \;.
\end{align}

We can also divide the sum into $c$ odd and $c$ even, to obtain:
\begin{align}
E_{Q,h}^{0,0}(\tau)
=\delta_{h\in \Lambda}&+\sum_{\substack{(c,d)=1\\d\text{ even}\\c>0}}\frac{\gamma_{Q,h}(c,d)}{(c\tau+d)^{p/2}(c\overline{\tau}+d)^{q/2}}
+\sum_{\substack{(c,d)=1\\d\text{ odd}\\c>0}}\frac{\gamma_{Q,h}(2c,2d)}{2^{(p+q)/2}(c\tau+d)^{p/2}(c\overline{\tau}+d)^{q/2}}\;.
\end{align}
The expression for $\gamma_{Q,h}(2c,2d)$ reads as
\begin{align}
\gamma_{Q,h}(2c,2d)=(2c)^{-(p+q)/2}e^{i\pi \sigma/4}\sum_{\ell \in \Lambda/(2c\Lambda)}e^{-\pi i dQ(\ell)/c}\;. 
\end{align}
If we now shift $\ell\to \ell+cx$, where $x$ is a basis vector with $Q(x)$ odd, then the summand is multiplied by $\exp(-i\pi dc)=(-1)^c$, where we assumed that $d$ is odd. Therefore for $c$ even we have $\gamma_{Q,h}(2c,2d)=2^{(p+q)/2}\gamma_{Q,h}(c,d)$, and for $c$ odd the sum vanishes. The answer then reduces to
\begin{align}
E_{Q,h}^{0,0}(\tau)&=\delta_{h\in \Lambda}+\sum_{\substack{(c,d)=1\\cd\in 2\mathbb{Z}\\c>0}}\frac{\gamma_{Q,h}(c,d)}{(c\tau+d)^{p/2}(c\overline{\tau}+d)^{q/2}} \;. \label{E00_sum}
\end{align}
This is the same expression for $E_{Q,h}$ \eqref{Siegel_Qh} for even $Q$, but with the additional constraint that $cd$ is even. 

Now that we have identified the Eisenstein series for the $(0,0)$ spin structure,
we can generate two more spin structures by the modular transformations (see Fig.~\ref{map}). 

Let us start with the case of $|\det Q|=1$ so that $h=0$. Then the other two Eisenstein series 
are derived from the modular transformations of the theta functions to be 
\begin{align}
E_{Q,0}^{0,1}(\tau)&=E_{Q,0}^{0,0}(\tau+1)\;,\\
E_{Q,0}^{1,0}(\tau)&=\frac{e^{i\pi\sigma/4}E_{Q,0}^{0,0}((\tau-1)/\tau)}{\tau^{p/2}\overline{\tau}^{q/2}}\;,
\end{align}
and consequently
\begin{align}
E_{Q,0}^{0,1}(\tau)&=1+\sum_{\substack{(c,d)=1\\c(d+1)\in 2\mathbb{Z}\\c>0}}\frac{\gamma_{Q,0}(c,d-c)}{(c\tau+d)^{p/2}(c\overline{\tau}+d)^{q/2}} \;,\label{E01_sum}\\
E_{Q,0}^{1,0}(\tau)&=\frac{e^{i\pi\sigma/4}}{\tau^{p/2}\overline{\tau}^{q/2}}+e^{i\pi\sigma/4}\sum_{\substack{(c,d)=1\\(c+1)d\in 2\mathbb{Z}\\d<0}}\frac{\gamma_{Q,0}(-d,c+d)}{(c\tau+d)^{p/2}(c\overline{\tau}+d)^{q/2}}\;. \label{E10_sum}
\end{align}
Now let us look at the $\gamma$'s that appear in the numerator of \eqref{E00_sum}, \eqref{E01_sum} and \eqref{E10_sum}. 
We expect that these expressions should coincide with the Chern-Simons invariants, as in the case of the even quadratic forms. We will see that this is indeed true in the next subsection. In this case, we have a spin Chern-Simons theory, and the phase of the one-loop determinant 
is given by the fermionic eta invariant, which can be written as a sum of the 
spin-independent eta invariant as well as the spin-dependent Arf invariant \cite{MR324709}. For $E_{Q,0}^{0,0}$ \eqref{E00_sum} $\gamma_{Q,h}(c,d)$ is simply the ordinary partition function with no spin structure. For $E_{Q,0}^{0,1}$ \eqref{E01_sum} we have 
\begin{align}
\gamma_{Q,0}(c,d-c)=c^{-(p+q)/2}e^{\pi i\sigma/4}\sum_{\ell\in \Lambda/c\Lambda}\exp(-\pi i dQ(\ell)/c)(-1)^{Q(\ell)} \;.
\end{align}
When $Q$ is rank $1$, this matches the nontrivial spin structure invariant in \cite{Okuda:2020fyl}. 

We can repeat similar computations for higher $|\det Q|$ 
by using the modular transformations,
\begin{align}
E_{Q,h}^{0,1}(\tau)&=e^{-2\pi i (q_W(h)-q_W(0))}E_{Q,h}^{0,0}(\tau+1)\;, \label{E01_formula}\\
E_{Q,h}^{1,0}(\tau)&=\frac{e^{i\pi\sigma/4}}{|\text{det }Q|^{1/2}\tau^{p/2}\overline{\tau}^{q/2}}\sum_{h'\in \Lambda^*/\Lambda}e^{2\pi i (Q(h,h')-q_W(h')+q_W(0))}E_{Q,h'}^{0,0}((\tau-1)/\tau)\;. \label{E10_formula}
\end{align}
Here we have defined
\begin{align}
    q_W(h): =\frac{1}{2}Q(h,h-W)+\frac{1}{8}Q(W,W) \in \mathbb{Q}/\mathbb{Z}\;.
\end{align}
Note that $q_W(h)$ does not depend on the choice of the representative from the quotient $\Lambda^*/\Lambda$,
since for $h\in \Lambda^*$ and $\ell\in \Lambda$
\begin{align}
    q_W(h+\ell) -q_W(h) =Q(h,\ell ) + \frac{Q(\ell) - Q(\ell, W)}{2} \in \mathbb{Z}\;,
\end{align}
as follows from the integrality of $Q$ and the definition of $W$.
The quadratic form $q_W$ on the discriminant $\mathscr{D}$ is a 
quadratic refinement of the bilinear form on $\mathscr{D}$ induced from $Q$:
\begin{align}
 q_W(h+h')-q_W(h)-q_W(h')+q_W(0)=Q(h, h')\;,
\end{align}
for $h, h'\in \mathscr{D}$.
Here on the right hand side $Q$ is regarded as a bilinear form on the discriminant $\mathscr{D}=\Lambda^*/\Lambda$ with values in $\mathbb{Q}/\mathbb{Z}$.

Notice that we can also compute these Eisenstein series by 
repeating the proof of the Siegel-Weil theorem in section \ref{sec.SW}.
 We can apply the same logic to the Eisenstein series for the remaining odd structure $(1,1)$,
and we find that $E^{1,1}_{Q,h}$ vanishes: this follows since
the theta function for the singlet is a modular form for some $\Gamma(N)$ with eigenvalue less than $\lambda_{\text{min}}$, and it is zero at all of the cusps. 

The three sums \eqref{E00_sum}, \eqref{E01_formula}, and \eqref{E10_formula} compute the Eisenstein series for each of the triplet of even spin structures under the modular group. 
We can again write these as sums over geometries, in terms of the corresponding invariants of spin Chern-Simons theories. In direct analogy to (\ref{eisensteininvariants}) we expect 
\begin{align}
\frac{E_{Q,h}^{\epsilon_1,\epsilon_2}(\tau)}{\eta(\tau)^{p}\overline{\eta}(\overline{\tau})^{q}}&=\sum_{g\in \Gamma_\infty\backslash\PSL(2,\mathbb{Z})}\frac{\langle \epsilon_1,\epsilon_2;h|\bm{U}(g^{-1})|0,0;h=0\rangle+\langle \epsilon_1,\epsilon_2;h|\bm{U}(g^{-1})|0,1;h=0\rangle}{\eta(g\cdot \tau)^{p}\overline\eta(g\cdot \overline{\tau})^{q}}\label{eisensteininvariantsspin}\;.
\end{align}
Here we have defined the states $|\epsilon_1\epsilon_2;h\rangle$ with spin structure $(\epsilon_1,\epsilon_2)$ and charge $h$. The logic is the same as in section \ref{sec.SW}: in order to obtain the behavior of the theta functions at an arbitrary cusp, we perform a modular transformation $g$ on the theta functions near the cusp at $\tau=i\infty$. The two spin structures $(0,0)$ and $(0,1)$ are the only spin structures for which the theta functions are nonvanishing at the cusp at $\tau=i\infty$, so the numerator of $(\ref{eisensteininvariantsspin})$ represents the transformation from $\tau=i\infty$ to an arbitrary cusp. \\
\indent In the next subsection we will check the formula (\ref{eisensteininvariantsspin}) in some specific examples. Before doing so, let us perform a preliminary consistency check. The formulas \eqref{E00_sum}, \eqref{E01_formula}, and \eqref{E10_formula} contain various congruence conditions on $c$ and $d$ modulo 2, which should be reproduced by \eqref{eisensteininvariantsspin}. To see how these conditions arise, note that the modular group acts on the triplet of even spin structures as the permutation group on three elements, as in figure \ref{map}. For example, if $c$ and $d$ are both odd, then up to framing ambiguities we have $g=STS$ modulo 2. This means that 
\begin{align}
    \langle 0,0;h|\bm{U}(g^{-1})|0,0,h=0\rangle=\langle 0,0;h|\bm{U}(g^{-1})|0,1;h=0\rangle=0 \;,
\end{align}
so that the summand of $E^{0,0}_{Q,h}$ vanishes. This is consistent with the constraint $cd\in 2\mathbb{Z}$ in \eqref{E00_sum}. The other congruence conditions follow in a similar manner. %

\subsection{Spin Chern-Simons Invariants}\label{scs}
 
In this subsection, we will show that spin Chern-Simons invariants appear in the novel Eisenstein series presented in the previous subsection, in a form consistent with (\ref{eisensteininvariantsspin}). We shall follow the approach of Jeffrey \cite{MR1175494}, who computed the Witten-Reshetikhin-Turaev (WRT) invariants \cite{Witten:1988hf,MR1091619} for non-spin Chern-Simons theory on lens spaces from  the action of $\PSL(2,\mathbb{Z})$ on the Hilbert space on a solid torus. Such an invariant is the partition function of the Chern-Simons theory, obtained as a matrix element of the gluing matrix $\bm{U}(g)$ with $g\in \PSL(2,\mathbb{Z})$ that glues two solid tori to give a lens space. 

In order to compute spin Chern-Simons invariants in an analogous manner, we shall employ the $\PSL(2,\mathbb{Z})$  transformations of elements of the Hilbert space  of a general Abelian \textit{spin} Chern-Simons theory on $\mathbb{T}^2\times \mathbb{R}$, which was derived explicitly by Belov and Moore \cite{Belov:2005ze} (with minor corrections in \cite{Stirling:2008bq}).\footnote{See \cite{Aghaei:2020otq} for analogous spin-structure-dependent computations of the matrix elements of the mapping class group action, in a different context of the analytic continuations of a supergroup Chern-Simons theory.}
Such a Hilbert space is labeled by the characteristic class $W\in \Lambda^*$ (where $\Lambda$ is the integral lattice characterizing the spin Chern-Simons theory) and a pair of spin structures, $\epsilon_1$ and $\epsilon_2$, defined on the boundary torus.
The matrix elements of an operator $\bm{O}$ acting from 
$H_{\epsilon_1, \epsilon_2,W}$ to $H_{\epsilon_1^\prime, \epsilon_2^\prime,W}$ are denoted
via the notation $\bm{O}_{h^\prime}^{~h}
\bigl [\begin{smallmatrix}\epsilon_1 && \epsilon_2\\
\epsilon_1^\prime && \epsilon_2^\prime\end{smallmatrix}\bigr ]$, where $h, h'$ label elements of the discriminant group $\mathscr{D}=\Lambda^*/\Lambda$.
In particular, the modular transformation operators are represented by matrices with the following elements (only nonzero elements are listed).\footnote{Our wavefunctions are the complex conjugates of those in \cite{Belov:2005ze,Stirling:2008bq}.} The $\bm{T}$ operator matrix elements are 
\begin{align}
\bm{T}_{h^\prime}^{~h}
\bigl [\begin{smallmatrix}0 && 0\\0 && 1\end{smallmatrix}\bigr ]&=
e^{-\frac{\pi i\sigma}{12}} e^{2\pi i[q_W(-h)-q_W(0)]}\delta_{h^\prime}^{~h} \;, \\
\bm{T}_{h^\prime}^{~h}
\bigl [\begin{smallmatrix}0 && 1\\0 && 0\end{smallmatrix}\bigr ]&=e^{-\frac{\pi i\sigma}{12}}
e^{2\pi i[q_W(h)-q_W(0)]}\delta_{h^\prime}^{~h}\;, \\
\bm{T}_{h^\prime}^{~h}
\bigl [\begin{smallmatrix}1 && 0\\1 && 0\end{smallmatrix}\bigr ]&=
\bm{T}_{h^\prime}^{~h}
\bigl [\begin{smallmatrix}1 && 1\\1 && 1\end{smallmatrix}\bigr ]=
e^{-\frac{\pi i\sigma}{12}}e^{2\pi iq_W(-h)}\delta_{h^\prime}^{~h}\;, 
\end{align}
while the $\bm{S}$ operator matrix elements are
\begin{align}
\bm{S}_{h^\prime}^{~h}
\bigl [\begin{smallmatrix}0 && 0\\0 && 0\end{smallmatrix}\bigr ]&=
\bm{S}_{h^\prime}^{~h}
\bigl [\begin{smallmatrix}1 && 0\\0 && 1\end{smallmatrix}\bigr ]=
|\det Q|^{-1/2}e^{-2\pi iQ(h',h)}\;, \\
\bm{S}_{h^\prime}^{~h}
\bigl [\begin{smallmatrix}0 && 1\\1 && 0\end{smallmatrix}\bigr ]&=
|\det Q|^{-1/2}e^{-2\pi iQ(h^\prime+{W},h)}\;, \\
\bm{S}_{h^\prime}^{~h}
\bigl [\begin{smallmatrix}1 && 1\\1 && 1\end{smallmatrix}\bigr ]&=
|\det Q|^{-1/2}e^{-2\pi iQ(h^\prime+{W},h)-4\pi i q_W(0)}\;.
\end{align}

To compute a spin Chern-Simons invariant, we shall concatenate these operators to form a gluing matrix, keeping in mind how each operator maps spin structures. 
In particular, the inverse of an operator maps spin structures in a direction opposite to that of the operator, e.g., the inverse of $\bm{S}_{h^{\prime}}^{~h}
\bigl [\begin{smallmatrix}1 && 0\\0 && 1\end{smallmatrix}\bigr ]$ is $(\bm{S}^{-1})_{h^{\prime}}^{~h}
\bigl [\begin{smallmatrix}0 && 1\\1 && 0\end{smallmatrix}\bigr ]$. The inverse $\bm{S}^{-1}$ can be computed with the help of an identity
\begin{equation} 
    |\det Q|^{-1}\sum_{h''\in \grp }e^{2\pi i Q(h-h',h'')}
    =\delta_{h,h'}\;.
\end{equation}

We shall first compute a spin Chern-Simons invariant for the lens space $L(c,\epsilon)$ (where $c>0$ and $\epsilon=\pm 1$) with trivial spin structure, and show that it takes a form that we expect from the Eisenstein series $E^{0,0}_{Q,h}$. The gluing matrix for this space is
\begin{equation}
    \bm{U}(g)=\left(
\begin{array}{cc}
\epsilon & 0\\
c & \epsilon 
\end{array}
\right)=\bm{S}^{\epsilon}\bm{T}^{-\epsilon c}\bm{S}^{-1}\;.
\end{equation}
The matrix element of interest (corresponding to a lens space with Wilson line insertion) is computed to be 
\begin{equation}
    \langle 00;h=0|\bm{S}^{\epsilon} \bm{T}^{-\epsilon c}\bm{S}^{-1}  |00;h\rangle =e^{\frac{\epsilon c i \pi \sigma}{12}}|\det Q|^{-1} \sum_{h'\in \grp}e^{-\epsilon c \pi i Q(h')}e^{2\pi i Q(h',h)}\;, 
\end{equation}
where $c$ \textit{must} be even to obtain a nonzero answer. This results in the $W$ dependence cancelling out due to the form of the $\bm{T}$ transformations. 
We now make use of the Gauss reciprocity formula
derived in \cite{MR2269836}, and described in appendix \ref{gauss}, which is 
\begin{equation}\label{gauss2}
 \sum_{h'\in \grp}e^{-\epsilon c \pi i Q(h')}e^{2\pi i Q(h',\Psi)}= e^{-\epsilon\pi i \frac{\sigma}{4}} |\textrm{det }Q|^{\frac{1}{2}} c^{-\frac{p+q}{2}} \sum_{\ell \in \Lambda/c\Lambda} e^{\frac{\epsilon}{ c}\pi i Q(\ell+\Psi)} 
\end{equation}
for $c$ even.
Using this formula, we obtain the spin Chern-Simons invariant
 \begin{equation}
    \langle 00;0|\bm{S}^{\epsilon} \bm{T}^{-\epsilon c}\bm{S}^{-1} |00;h \rangle=e^{\frac{\epsilon c\pi i\sigma}{12}}e^{-\epsilon\pi i \frac{\sigma}{4}}  \dt c^{-\frac{p+q}{2}} \sum_{\ell \in \Lambda/c\Lambda} e^{\frac{\epsilon}{c}\pi i Q(\ell + h)}\;, 
\end{equation}
where $c$ is restricted to be even. This is the complex conjugate of the expression $\gamma_{Q,h}(c,d)$ that appears in the Eisenstein series $E^{0,0}_{Q, h}$ (c.f. \eqref{E00_sum}), for $d=\epsilon$, multiplied by the complex conjugate of the overall phase in (\ref{ugammamatch}). This formula generalizes the result of Okuda et al.\ \cite{Okuda:2020fyl} involving a single $\U(1)$ gauge group.
 
Next, we shall show that a spin Chern-Simons invariant appears in $E^{0,1}_{Q,h}$. To this end we shall compute a spin Chern-Simons invariant for the lens space $L(c,\epsilon)$ with nontrivial spin structure, i.e., we would like to compute 
\begin{equation}
        \langle 01;0|\bm{S}^{\epsilon} \bm{T}^{-\epsilon c}\bm{S}^{-1}|01;h \rangle \;.
\end{equation}
Here, the concatenation of operators begins on the right with $(\bm{S}^{-1})_{h^\prime}^{~h}
\bigl [\begin{smallmatrix}0 && 1\\1 && 0\end{smallmatrix}\bigr ]$. Since $\bm{T}_{h^\prime}^{~h}
\bigl [\begin{smallmatrix}1 && 0\\1 && 0\end{smallmatrix}\bigr ]$ is nonzero, there is no restriction on $c$. Furthermore, for $\epsilon = -1$, we must end the concatenation with
 \begin{equation}\label{swi}
     (\bm{S}^{-1})_{h^\prime}^{~h}
\bigl [\begin{smallmatrix}1 && 0\\0 && 1\end{smallmatrix}\bigr ]= |\det Q|^{-1/2}e^{2\pi iQ(h^\prime,h +{W})} \;.
 \end{equation}
 Then, we obtain 
 \begin{equation}\label{lal}
    \langle 01;0|\bm{S}^{\epsilon} \bm{T}^{-\epsilon c}\bm{S}^{-1} |01;h \rangle =e^{\frac{\epsilon c\pi i\sigma}{12}}|\det Q|^{-1} \sum_{h'\in \grp}e^{-2\epsilon c \pi i q_W(-h')}e^{2\pi i Q(h',h)} \;.
\end{equation}
We can write the right hand side of \eqref{lal} as
\begin{equation}
\begin{aligned}
    &e^{\frac{\epsilon c\pi i\sigma}{12}}|\det Q|^{-1} \sum_{h'\in \grp}e^{-\epsilon c \pi i \big(Q(h') + Q(h',W)+\frac{Q(W,W)}{4}\big)}e^{2\pi i Q(h',h)}\\
   & = e^{\frac{\epsilon c\pi i\sigma}{12}} e^{-\epsilon c \pi i\frac{Q(W,W)}{4}} |\det Q|^{-1}\sum_{h'\in \grp}e^{ -\epsilon c\pi i Q(h')}e^{ 2\pi i Q(h',h-\frac{\epsilon c}{2} W)}   \;. \end{aligned}
\end{equation}
To obtain a familiar expression from this we ought to use a Gauss reciprocity formula.\footnote{See \cite{Ganor:2019nnv} for another discussion of reciprocity formulas in the study of Abelian Chern-Simons theories.} 
For $c$ even, we can use \eqref{gauss2} (with $\Psi=h-\frac{\epsilon c}{2}W$), while for $c$ odd, we ought to use the more general formula
\begin{equation}\label{reciprocity2}
\begin{aligned}
    &\frac{1}{\sqrt{|\textrm{det }Q|}}\sum_{h'\in \Lambda^*/\Lambda}e^{-\pi i \epsilon c Q(h',h')}e^{2\pi i  Q(h',h-\frac{\epsilon c}{2}W)}
    =c^{-\frac{p+q}{2}}e^{-\frac{\pi i \sigma(Q)\epsilon }{4}}\sum_{\ell \in \Lambda/c\Lambda} e^{\frac{\pi i \epsilon}{c}Q(\ell+h-\frac{\epsilon  cW}{2})} \;,
    \end{aligned}
\end{equation}
described in appendix \ref{gauss}.
Doing so,
we find the right hand side of \eqref{lal} to be 
\begin{equation}
\begin{aligned}
 & e^{\frac{\epsilon c\pi i\sigma}{12}} e^{-\epsilon c \pi i\frac{Q(W,W)}{4}}  e^{-\epsilon\pi i \frac{\sigma}{4}}  \dt c^{-\frac{p+q}{2}} \sum_{\ell \in \Lambda/c\Lambda} e^{\frac{\epsilon}{c}\pi i Q(\ell+h-\frac{\epsilon c}{2} W)}
\\&= e^{\frac{\epsilon c\pi i\sigma}{12}}
    e^{-\epsilon \pi i \frac{\sigma}{4}}  \dt c^{-\frac{p+q}{2}}\sum_{\ell\in \Lambda/ c \Lambda} e^{\frac{\epsilon}{c}\pi i Q(\ell + h)}e^{-i\pi Q(\ell+h,W)} \;.
\end{aligned}
\end{equation}
(For $h=0$ and $\Lambda$ a rank 1 lattice, this agrees exactly with the result of Okuda et al.\ \cite{Okuda:2020fyl}.)
Up to the overall phase in (\ref{ugammamatch}), this in fact takes the form of the complex conjugate of $\gamma_{Q,h}(c,d-c)$ that appears in $E^{0,1}_{Q,h}$, i.e.,
\begin{equation}
E_{Q,h}^{0,1}(\tau)=\delta_{h\in \Lambda} +\sum_{\substack{(c,d)=1, \\c (d+1)\in 2\mathbb{Z}, \\c>0}}\frac{c^{-(p+q)/2}\dt e^{\pi i\sigma/4}\sum_{\ell\in \Lambda/c\Lambda}e^{-\pi i \frac{d}{c}Q(\ell+h)}e^{\pi i Q(\ell+h,W)} }{(c\tau+d)^{p/2}(c\overline{\tau}+d)^{q/2}} \;. 
\end{equation}
\indent Finally, let us consider $E^{1,0}_{Q,h}$, for which the relevant matrix element is 
\begin{align}
\langle 00;0|\bm{S}^\epsilon \bm{T}^{-\epsilon c}\bm{S}^{-1}|10;h\rangle=\frac{e^{\frac{\epsilon c\pi i \sigma}{12} }}{|\det Q|}\sum_{h'\in \Lambda^*/\Lambda}e^{-2\pi i Q(h',h)-\pi ic\epsilon Q(h')-\pi i Q(W,h')} \;,\label{e10matrix}
\end{align}
where $c$ is restricted to be odd. To match this to the Eisenstein series, we expand (\ref{E10_formula}),
\begin{align}
\begin{split}
E^{1,0}_{Q,h}(\tau)&=\frac{e^{i\pi\sigma/4}}{|\text{det }Q|^{1/2}\tau^{p/2}\overline{\tau}^{q/2}}\\
&\qquad +\frac{e^{i\pi \sigma/4}}{|\det Q|^{1/2}}\sum_{h'\in \Lambda^*/\Lambda}\sum_{\substack{(c,d)=1 \\(c+1)d\in 2\mathbb{Z} \\d<0}}e^{2\pi i (Q(h,h')-q_W(h')+q_W(0))}\frac{\gamma_{Q,h'}(-d,c+d)}{(c\tau+d)^{p/2}(c\overline{\tau}+d)^{q/2}} \;.
\end{split}
\end{align}
For $\epsilon=d=-1$, we have
\begin{align}
e^{2\pi i (Q(h,h')-q_W(h')+q_W(0))}\gamma_{Q,h'}(-d,c+d)=e^{\pi i\sigma/4} |\textrm{det} \, Q|^{-\frac{1}{2}}  e^{2\pi i Q(h,h')+\pi iQ(h',W)-\pi i cQ(h')} \;,
\end{align}
and complex conjugating indeed gives (\ref{e10matrix}) up to the overall phase. 

\section{Holographic Dual before Ensemble Average}\label{sec.before}

We have seen that the CFT partition function gives that of the Abelian spin Chern-Simons theory 
after the ensemble average. However, we do not encounter such an ensemble average
in the standard discussion of holography. One might therefore wonder
if there exists a holography \emph{before} ensemble average,
so that the ensemble average of this more ``fine-grained'' holography
gives the holography of the previous sections as a ``coarse-grained'' counterpart
after the ensemble average (see also \cite{Eberhardt:2021jvj}). 
In this section, we address this question.

In order to discuss holography before ensemble averages, 
we need to incorporate the dependence on the moduli $\mathcal{M}_Q$.
Since a point of the moduli space 
gives a decomposition into left movers and right movers (recall from section \ref{sec.lattice}),
one possibility is to perform such a decomposition to the Chern-Simons theory, so that we have an action
\begin{align}
S_{\rm CS}
&=i\sum_{i,j=1}^{p} \frac{(Q_L)_{ij }}{8\pi} \int_M A_L^i \wedge d A_L^j -i \sum_{i,j=1}^{q} \frac{(Q_R)_{ij }}{8\pi} \int_M A_R^i \wedge d A_R^j\;,
\label{CS_pm0}
\end{align}
where the ``left-moving'' (resp.\ ``right-moving'') gauge fields $A_L^{1, \dots, p}$ (resp.\ $A_R^{1, \dots, q}$)
are linear combinations of the gauge fields $A^{1, \dots, p+q}$.

Unfortunately, this does not quite work as it is, since 
such a linear transformation among the gauge fields is in general not compatible 
with the quantization conditions for the gauge fields (the gauge groups are $\U(1)$, not $\mathbb{R}$).
This is related to the fact that the boundary of the Chern-Simons theory 
always gives a rational CFT, while the boundary CFT is irrational at a generic point 
of the CFT moduli space.

In \cite{Gukov:2004id} it was recognized that we can realize irrational CFTs on the boundary
if we instead consider a Maxwell-Chern-Simons theory \cite{Deser:1981wh,Deser:1982vy}. 
The theory is defined by the action 
\begin{equation}\label{mym}
   S_{\rm MCS}= \frac{1}{16\pi^2}\sum_{i,j=1}^{p+q} \int_M\bigg(-\frac{1}{2e^2}\lambda_{ij}^{-1} d A^i\wedge * d A^j + 2\pi i Q_{ij}A^i \wedge d A^j \bigg) \;,
\end{equation}
where $e^2$ is the coupling which has dimensions of mass, and $\lambda^{-1}$ is a dimensionless, symmetric, positive definite matrix with determinant one. Since $e^2$ is dimensionful, the Yang-Mills term is 
irrelevant and the Chern-Simons term is expected to dominate in the IR. This gives the 
topological limit $e^2\rightarrow \infty$, leaving only the Chern-Simons term.
The effect of the Yang-Mills term, however, does not quite 
go away, since the quantization conditions for the gauge fields in the topological limit depend on 
the parameters $\lambda$, and hence on a point of the moduli space $\mathcal{M}_Q$.

We will see that in this setup we can identify  a duality between the resulting Abelian Chern-Simons theory defined with respect to the quadratic form $Q$, and a CFT associated with an integral lattice with the same quadratic form, at each point of the moduli space of the latter. 
In contrast to the previous discussion that the exotic bulk theories after ensemble averages are only approximately Chern-Simons theories, 
here we find that the bulk theory is given precisely by the Chern-Simons theories.
In order to simplify the discussion we restrict to the case of the trivial spin structure.

The aforementioned quantization of the Maxwell-Chern-Simons theory is performed on the infinite volume limit of the solid torus, i.e., $M=T^2\times \mathbb{R}$. Picking a complex structure $\tau$ on the torus with flat metric, the basis of  wavefunctions for the topological sector of the theory was shown in \cite{Gukov:2004id,Belov:2005ze} to be 
\begin{equation}
    \Psi_{h}=|\textrm{det }Q|^{\frac{1}{4}}e^{-\frac{1}{8\pi} \int_{T^2}\sum_{i,j}\mu_{ij} A^i \wedge *A^j}\frac{\vartheta_{Q,h}(\tau,\xi(A))}{\eta(\tau)^p \overline{\eta}(\overline{\tau})^q} \;,
\end{equation}
for $h\in \Lambda^*/\Lambda$, where $\mu$ is a symmetric positive definite matrix defined below, and where the theta function is defined as
\begin{equation}\label{cstheta}
    \vartheta_{Q,h}(\tau,\xi(A)) := e^{\frac{\pi}{2\, \textrm{Im}\, \tau}(Q_L(\xi)+Q_R(\xi))}\sum_{\ell\in\Lambda}e^{i\pi\tau Q_L(\ell+h)-i\pi\overline{\tau}Q_R(\ell+h) +2\pi i Q(\ell+h,\xi)} \;,
\end{equation}
with $Q_L:=\frac{1}{2}(Q+\mu)$, $Q_R:=\frac{1}{2}(-Q+\mu)$, and $\xi(A):=-\frac{1}{\sqrt{2\pi}} (P_-(i\imt A_{\zbar});P_+(i\imt A_{z}))$, where $P_{\pm}:=\frac{1}{2}(1\pm \mu^{-1} Q)$ are projection operators onto left/right movers. The matrix $\mu$ takes the form
\begin{equation}
\mu = \lambda^{-1/2} \mathcal{O} \begin{pmatrix} \Delta^+ & 0 \\ 0 & -\Delta^-\\\end{pmatrix}  \mathcal{O}^{T} \lambda^{-1/2} \;,
\end{equation}
where $\Delta^{\pm}$ are diagonal matrices satisfying $\Delta^+_{ii}>0$ and $\Delta^-_{ii}<0$, and $\mathcal{O}$ is a real orthogonal matrix that diagonalizes $\lambda^{1/2}Q \lambda^{1/2}$ such that 
\begin{equation}
Q = \lambda^{-1/2} \mathcal{O} \begin{pmatrix} \Delta^+ & 0 \\ 0 & \Delta^-\\\end{pmatrix} \mathcal{O}^{T} \lambda^{-1/2}\;.
\end{equation}
We thus observe that the moduli that enter the theta function in \eqref{cstheta} arise from the matrix $\lambda$ whose elements enter the kinetic term in \eqref{mym}. 

To elucidate the duality with a CFT before averaging, let us specialize to the case of gauge group $\U(1)\times \U(1)$ with gauge fields denoted $A$ and $B$, where $e^2= e_Ae_B$ and 
\begin{equation}
    Q=\begin{pmatrix}
0 & \frac{k}{2} \\
\frac{k}{2} & 0
\end{pmatrix}
\;,
\quad
\lambda=    \begin{pmatrix}
\frac{e_A}{e_B} & 0 \\
0 & \frac{e_B}{e_A}
\end{pmatrix} \;.
\end{equation}
In this case,
\begin{equation}
\mu=\frac{k}{2}\begin{pmatrix}
\frac{e_B}{e_A} & 0 \\
0 & \frac{e_A}{e_B}
\end{pmatrix}. 
\end{equation}
Now, the vector $\xi(A)=-\frac{1}{\sqrt{2\pi}} (P_-(i\imt A_{\zbar});P_+(i\imt A_{z}))$ involves components of the fields $P_-A:=A^{-}$ and $P_+A:=A^{+}$. If we 
were to compute the path integral of this Maxwell-Chern-Simons theory on the solid torus, we ought to obtain a state in the Hilbert space of the theory as a function of the boundary values of the fields. If we were to choose the boundary conditions $A_{\zbar}^-=0$ and $A_{z}^+=0$, we find that the basis of wavefunctions simplifies to 
\begin{equation}\label{swf}
      \Psi_{h}=\sqrt{k} \, \frac{\vartheta_{Q,h}(\tau,0)}{|\eta(\tau)|^2} \;,
\end{equation}
which is, up to a factor of $\sqrt{k}$, the CFT partition function \eqref{ZQh} for $p=q=1$, once we identify $\mu$ with the Hamiltonian $H$.
It is in this sense that there is a duality between the topological sector of Maxwell-Chern-Simons theory and the CFTs studied in  section 2.2. 

Note that, since the Chern-Simons sector of the action can be recast as $S_{\rm CS}=\frac{i}{16 \pi} k\int (A^{+}\wedge dA^{+} -A^{-}\wedge dA^{-})$, the boundary conditions $A_{\zbar}^-=0$ and $A_{z}^+=0$ resemble the boundary conditions (in Euclidean signature) used by Coussaert, Henneaux and Van Driel \cite{Coussaert:1995zp} in relating $SL(2,\mathbb{R})\times SL(2,\mathbb{R})$ Chern-Simons theory to the $SL(2,\mathbb{R})$ WZW model. This would indeed be the case if the ratio $e_B/e_A$ was rational, but not otherwise since $A^+$ and $A^-$ cannot then be defined as nontrivial connections, and would not be truly independent.

For general $Q$ and $\lambda$, the boundary conditions $P_-A_{\zbar}=0$ and $P_+A_{z}=0$ lead us to the same conclusion of a duality before averaging between the topological limit of Maxwell-Chern-Simons theory and the aforementioned CFTs for each value of their moduli. These observations, in fact, generalize to the case of spin Maxwell-Chern-Simons theory on higher genus handlebodies with nontrivial spin structure, whose wavefunctions were derived in \cite{Belov:2005ze}. The CFT dual to this theory for genus one was studied in section 3.

The discussion up to this point makes clear
one can formulate the holography not only for the partition functions but also 
at the level of states inside the Hilbert space. To see this, first note that the
CFT partition function on the two-torus should be regarded as a wavefunction of the holographic dual on the Hilbert space $\mathcal{H}(\mathbb{T}^2)$ associated with the two-torus.
In order to better represent this fact, we introduce a new bra-ket notation
\begin{align}
 Z_{Q,h}(\tau, \taubar; m)  \rightsquigarrow |  h;m  \rangle \;,
\end{align}
so that the Hilbert space $\mathcal{H}(\mathbb{T}^2)$  is spanned by $|  h;m  \rangle $ with $h\in \mathscr{D}$.
Note that  we are making the dependence on $Q$ and $\tau, \taubar$ to be implicit.
In this notation, the modular transformation rules of the CFT partition functions \eqref{CFT_Z_TS} coincide with 
the transformation rules \eqref{CS_TS} in the holographic dual.
In other words, both bulk and the boundary give exactly the same pair $(\mathcal{H}, \mathcal{R})$ of the Hilbert space $\mathcal{H}(\mathbb{T}^2)$
and a representation $\mathcal{R}$ of the mapping class group
$\PSL(2, \mathbb{Z})$ on the Hilbert space. Moreover, such a pair is independent of the
CFT moduli space, and hence is preserved by the ensemble average.

It is interesting to note that different choices of the quadratic form can generate equivalent
representations of the mapping class group, as discussed in \cite{Belov:2005ze}.
Such a equivalence is constrained more strongly in our case, since 
in our partition functions we have dependence on both the rank $p+q$ and the signature $\sigma=p-q$ of the quadratic form,
while for \eqref{CS_TS} only $p-q$ modulo $24$ enters into the representation of the mapping class group.
There is no inconsistency in these statements, since two Abelian Chern-Simons theories
which are equivalent in the sense of \cite{Belov:2005ze} can still lead to 
different partition functions when we consider manifolds with boundary, with different boundary conditions imposed.

\section{Discussion}\label{sec.discussion}

One of the interesting findings in the analysis of the holographic duality in this paper is that
once we have an ensemble average over the CFT moduli space then the sum over geometries in the bulk
is \emph{automatically} incorporated. 
We propose that this is a general lesson for holographies involving ensemble averages. If true, this can have far-reaching consequences in quantum gravity---instead of
summing over geometries (as you would do in theories of quantum gravity)
one can consider the ensemble averages of dual CFTs!

This duality between the CFT moduli space
and the moduli space of Riemann surfaces is closely related with the mathematical concept of Howe duality
and reductive dual pairs \cite{MR463359,MR986027,MR724015,MR1005856}: the symmetry of the CFT moduli space $G=\rmO(p,q; \mathbb{R})$
and the mapping class group for the spacetime surface $H=\Sp(2g; \mathbb{R})$
are embedded inside a larger symplectic group $G^\sharp=\Sp(2g(p+q);\mathbb{R})$ (or rather its double cover, the metaplectic group),
and are mutual centralizers inside it. Moreover, the Weil representation of the $G^\sharp$
are decomposed into irreducible components of $G$ and of $H$, where there exists a one-to-one correspondence
between those of $G$ and of $H$. The Siegel-Weil formula can be regarded as a reflection of a more general statement on 
modular forms of $G$ and $H$, known as the theta correspondence. It would be interesting to see if
such mathematical discussion sheds further light on the discussions of holography and quantum gravity.

\section*{Acknowledgements}

This work grew out of a study group held at Kavli IPMU, and we would like to thank other participants 
for taking part in the stimulating discussions. 
We would also like to thank Yuto Moriwaki, Takuya Okuda and Hirosi Ooguri for discussions related to this work. 
This research was supported in part by WPI Research Center Initiative, MEXT, Japan. 
This research was also supported by the JSPS Grant-in-Aid for Scientific Research (20K14465 [MD], 19H00689 [MA and MY], 17KK0087, 19K03820, 20H05850 and 20H05860 [MY]), the UC Berkeley Center for Japanese Studies under the CJS fellowship [JML], the U.S. Department of Energy under Contract DE-AC02-05CH11231 [JML], and the National Science Foundation under grant PHY-1316783 [JML].

\begin{appendix}
%
\section{Differential Equation for Theta Functions}\label{diffeqappendix}

In this appendix, we derive the differential equation satisfied by the theta functions discussed above. 
To be concrete, we consider $p\geq q$. In the derivation of this differential equation, following \cite{Obers:1999um}, we shall assume that the quadratic form is written in terms of a $(p+q)$-dimensional matrix of the form \begin{align}Q = \begin{pmatrix} 
& & \mathbb{I}_{q,q} \\ 
 & \mathbb{I}_{p-q,p-q} & \\
\mathbb I_{q,q} & &
 \end{pmatrix}.
 \end{align} However, the resulting differential equation should also hold for generic quadratic forms since they are obtained by conjugation with an element of $GL(p+q; \mathbb{R})$. The action of the quadratic form $H$, which is the majorant of $Q$, on a lattice point $\ell= (n_i,l_M,w^i)$ is expressed in terms of moduli $\{G_{ij},B_{ij}, A_{iM}\}$ as $H(\ell)=\ell^T\mathcal{G}\ell$, where
\begin{align}
\mathcal{G} = \begin{pmatrix}
		G^{ij} & -G^{ik}A_{k}^N & -G^{ik}W_{kj}\\
		-G^{jk}A_{k}^M & \delta^{MN} + G^{mn}A_{m}^N A_{n}^M & A_{j}^M+G^{mn}A_{m}^MW_{nj}\\
		-W^T_{ik}G^{kj} & A_{i}^N+G^{mn}A_{m}^NW^T_{in} & G_{ij}+A_{iB}A_{j}^B + G^{mn}W^T_{in}W_{mj}
\end{pmatrix}.
\end{align}
Here we have defined $W_{ij} = B_{ij}+\frac{1}{2}A_{iM}A_{j}^M$. The lowercase Latin indices take on values $\{1,...,q\}$ while the uppercase Latin indices take on values $\{1,...,p-q\}$. The target space metric moduli are symmetric $G_{ij} = G_{ji}$ and the 2-form field moduli are anti-symmetric $B_{ij} = -B_{ji}$. We follow the same procedure as~\cite{Obers:1999um} and obtain the metric on $\mathcal{M}_{Q}$ using $ds^2 = -\frac{1}{2}\text{Tr}(d\mathcal{G}d\mathcal{G}^{-1})$, which gives
\begin{align}\label{coset_metric}
    ds^2 & = G^{ij}G^{mn}(dG_{im}dG_{jn}+dB_{im}dB_{jn}) + 2G^{ij}dA_{iM}dA_{j}^M+2G^{ij}G^{mn}A_{iM}dA_{m}^MdB_{jn}\nonumber\\
            &\quad +\frac{1}{2}G^{ij}G^{mn}\bigg(A_{iM}A_{jN}dA_{m}^NdA_{n}^M - A_{iM}A_{mN}dA_{n}^MdA_{j}^N \bigg) \;.
\end{align}
A straightforward calculation yields the Laplacian on $\mathcal{M}_{Q}$
\begin{align}
\label{eq:genlap}
    \Delta_{\mathcal{M}_Q} &= \frac{1}{4}G_{ms}G_{nt}(\partial_{\tilde{G}_{mn}}\partial_{\tilde{G}_{st}}+\partial_{B_{mn}}\partial_{B_{st}})	+\frac{1}{2}\bigg(1-\frac{ p-q}{2}\bigg)G_{mn}\partial_{\tilde{G}_{mn}} + \frac{\delta_{MN}G_{mn}}{2}\partial_{m M}\partial_{n N} \;,
\end{align}
where
\begin{align}
\partial_{m,M}\partial_{n,N} =\bigg(\partial_{A_{mM}} + \frac{1}{2}A_{k}^M\partial_{B_{mk}}\bigg)
\bigg(\partial_{A_{nN}} + \frac{1}{2}A_{j}^N\partial_{B_{nj}}\bigg)\;.			
\end{align}
Here we have introduced the diagonally rescaled metric $\tilde{G}_{ij} = (1-\delta_{ij}/2)G_{ij}$ so that the derivatives above act as
\begin{align}
        \partial_{\tilde{G}_{st}}G_{mn} &= \delta^s_m\delta^t_n + \delta^t_m\delta^s_n \;,\\
        \partial_{B_{st}}B_{mn} &= \delta^s_m\delta^t_n - \delta^t_m\delta^s_n \;.
\end{align}
With a bit of effort, it is possible to show that the theta functions above satisfy the differential equation 
\begin{align}
\label{eq:generalizedlaplacian}
			\bigg(-\tau_2^2(&\partial_{2}^2+\partial_{1}^2)	-\frac{(p+q)\tau_2}{2}\partial_{2}-i\frac{ (q-p)\tau_2}{2}\partial_{1}+\Delta_{\mathcal{M}_Q}		\bigg)\vartheta_{Q,h}(\tau, \taubar;m) =0 \;.
\end{align}
Note that $\vartheta_{Q,h}(\tau,\taubar;m)$ stands for theta functions with or without spin structure. Furthermore, upon averaging, the moduli-dependent Laplacian drops out so that the averaged theta function satisfies the same differential equation as the related Eisenstein series. 

For illustration, we can consider the simplest example of $(p,q) = (2,1)$. Then \eqref{eq:generalizedlaplacian} simplifies to
\begin{align}
    \bigg(\tau_2^2(\partial_1^2 + \partial_2^2)+\frac{3\tau_2}{2}\partial_2 -\frac{i\tau_2}{2}\partial_1 -\frac{R^2}{4}\partial_R^2 - \frac{R^2}{2}\partial^2_A \bigg)\vartheta_{Q} = 0 \;.
\end{align}
\section{Theta Functions for Odd Lattices}\label{sec.oddtheta}

The theta functions for a given spin structure are not linearly independent. They satisfy the charge conjugation relations
\begin{align}
\label{conjrelationsthetaodd}
\begin{split}
    \vartheta_{0,0,h}(\tau)&=\vartheta_{0,0,-h}(\tau) \;,\\
    \vartheta_{1,0,h}(\tau)&=\vartheta_{1,0,-W-h}(\tau)\;,\\
    \vartheta_{0,1,h}(\tau)&=e^{2\pi i Q(h,W)}\vartheta_{0,1,-h}\;,\\
    \vartheta_{1,1,h}(\tau)&=e^{2\pi i Q(h+W/2,W)}\vartheta_{1,1,-W-h}(\tau)\;.
    \end{split}
    \end{align}
The modular transformations are
\begin{align}
\label{modtrafothetaodd}
\begin{split}
\vartheta_{0,0,h}(\tau+1)&=\exp(2\pi i (q_W(h)-q_W(0)))\,\vartheta_{0,1,h}(\tau)\;,\\
\vartheta_{0,1,h}(\tau+1)&=\exp(2\pi i (q_W(-h)-q_W(0)))\,\vartheta_{0,0,h}(\tau)\;,\\
\vartheta_{1,0,h}(\tau+1)&=\exp(2\pi i q_W(-h)))\,\vartheta_{1,0,h}(\tau)\;,\\
\vartheta_{1,1,h}(\tau+1)&=\exp(2\pi i q_W(-h)))\,\vartheta_{1,1,h}(\tau)\;,\\
\vartheta_{0,0,h}(-1/\tau)&=\frac{e^{-i\pi \sigma/4}\tau^{p/2}\overline{\tau}^{q/2}}{\sqrt{|\det Q|}}\sum_{h'}e^{-2\pi i Q(h,h')}\,\vartheta_{0,0,h}(\tau)\;,\\
\vartheta_{0,1,h}(-1/\tau)&=\frac{e^{-i\pi \sigma/4}\tau^{p/2}\overline{\tau}^{q/2}}{\sqrt{|\det Q|}}\sum_{h'}e^{-2\pi i Q(h,h')}\,\vartheta_{1,0,h}(\tau)\;,\\
\vartheta_{1,0,h}(-1/\tau)&=\frac{e^{-i\pi \sigma/4}\tau^{p/2}\overline{\tau}^{q/2}}{\sqrt{|\det Q|}}\sum_{h'}e^{2\pi i Q(h,h')}\,\vartheta_{0,1,h'}(\tau)\;,\\
\vartheta_{1,1,h}(-1/\tau)&=\frac{e^{-i\pi \sigma/4-\pi i(W,W)/2}\tau^{p/2}\overline{\tau}^{q/2}}{\sqrt{|\det Q|}}\sum_{h'}e^{-2\pi i Q(h+W,h')}\,\vartheta_{1,1,h'}(\tau)\;.
\end{split}
\end{align}
These transformation rules can be checked explicitly from the definitions with the help of Poisson resummation.

\section{Gauss Reciprocity Formulas}\label{gauss}

In this appendix, we collect Gauss reciprocity formulas that we use in the main text, based on the results of Deloup and Turaev \cite{MR2269836}. 

Consider a pair of lattices $\Lambda$ and $\Lambda'$, defined with the quadratic forms $Q$ and $Q'$ respectively. 
Let us consider a tensor product $\Lambda \otimes \Lambda'$ (over $\mathbb{Z}$) with a quadratic form $\widehat{Q}:=Q \otimes Q'$,
as well as a Wu class $z$ on it.
Theorem 2 of \cite{MR2269836} can then be stated as 
\begin{equation}\label{DT_Thm_2}
    \frac{1}{\sqrt{|A|}}\sum_{x\in A}e^{{\pi i }[\widehat{Q}(x,x)-\widehat{Q}(x,z)]}
      =\frac{1}{\sqrt{|B|}}e^{\frac{\pi i \sigma(Q)\sigma(Q')}{4}}\sum_{y\in B} e^{-\pi i \widehat{Q}(y-\frac{z}{2},y-\frac{z}{2})} \;,
\end{equation}
where  $A:=(\Lambda^*/\Lambda)\otimes \Lambda'$ and $B:=\Lambda\otimes (\Lambda'^*/\Lambda')$.

Now, pick $\Lambda'=c\Z$, where $c>0$ and odd, and $Q'=1/(\epsilon c)$ where $\epsilon =\pm 1$. 
The Wu class $W'\in \Lambda'^*/2\Lambda'^*$ on $\Lambda'$ is then given by $W'=\epsilon c$, since $\frac{1}{\epsilon c}W'x=\frac{1}{\epsilon c}x^2$ for any $x\in c\Z$. Let us choose a Wu class $W\in \Lambda^*/2\Lambda^*$ on $\Lambda$,
and set $z:=-W\otimes W'-2\Psi \otimes 1$ with $\Psi \in \Lambda^*$. We can verify that $z$ is a Wu class on the tensor product $\Lambda \otimes \Lambda'$.
Since we have $A=(\Lambda^*/\Lambda)\otimes c \simeq \Lambda^*/\Lambda$ and $B=\Lambda\otimes (\mathbb{Z}/c \mathbb{Z}) \simeq \Lambda/(c\Lambda)$, \eqref{DT_Thm_2} takes the form 
\begin{equation}
\begin{aligned}
    \frac{1}{\sqrt{|\Lambda^*/\Lambda|}}\sum_{h'\in \Lambda^*/\Lambda}e^{\pi i [\widehat{Q}(h'\otimes \epsilon c, h'\otimes \epsilon c)-\widehat{Q}(h'\otimes \epsilon c,z)]}
        &=\frac{1}{\sqrt{|\Lambda/c\Lambda|}}e^{\frac{\pi i \sigma(Q)\epsilon }{4}}\sum_{\ell \in \Lambda/c\Lambda} e^{-\pi i [\widehat{Q}(\ell\otimes 1-\frac{z}{2}, \ell \otimes 1-\frac{z}{2})]} \;.
         \end{aligned}
\end{equation}
Using $z=(-\epsilon c W-2\Psi) \otimes 1$ and $\widehat{Q}=Q\otimes Q'$ with $Q'=1/(\epsilon c)$,
this can be rewritten as
\begin{equation}\label{reciprocity2_app}
\begin{aligned}
    &\frac{1}{\sqrt{|\textrm{det }Q|}}\sum_{h'\in \Lambda^*/\Lambda}e^{\pi i \epsilon c Q(h',h')}e^{2\pi i  Q(h',\frac{\epsilon c}{2}W+{\Psi})}
    =c^{-\frac{p+q}{2}}e^{\frac{\pi i \sigma(Q)\epsilon }{4}}\sum_{\ell \in \Lambda/c\Lambda} e^{-\frac{\pi i }{\epsilon  c}Q(\ell+\frac{\epsilon  cW}{2}+\Psi)} \;.
    \end{aligned}
\end{equation}
This is precisely the formula we use in Section \ref{scs}. Note that for $c$ even, we have $W'=0$ and the dependence on $W$ is no longer present, with $z=-2\Psi \otimes 1$. 
\end{appendix}
\bibliographystyle{nb}
\bibliography{narain}


\end{document}